\documentclass[preprint2]{aastex63}

\usepackage{subfigure}
\usepackage{url}
\usepackage{hyperref}
\usepackage{epsfig}
\usepackage{graphicx}
\usepackage{sidecap}
\usepackage{hanging}
\usepackage{color}
\usepackage{multirow}
\usepackage{enumerate}
\usepackage{natbib}
\usepackage{amssymb}
\usepackage{amsmath}
\usepackage[bottom]{footmisc}
\usepackage{apjfonts}

\newcommand{\sci}{Science}
\newcommand{\jatis}{JATIS}

\newcommand{\kepler}{{\it Kepler}}
\newcommand{\tess}{{TESS}}

\providecommand{\e}[1]{\ensuremath{\, \times \, 10^{#1}}}

\newcommand{\planet}{{TOI-2180~b}}
\newcommand{\host}{{TOI-2180}}

\providecommand{\bjdtdb}{\ensuremath{\rm {BJD_{TDB}}}}

\providecommand{\msun}{\ensuremath{\,M_\Sun}}
\providecommand{\rsun}{\ensuremath{\,R_\Sun}}
\providecommand{\lsun}{\ensuremath{\,L_\Sun}}
\providecommand{\mj}{\ensuremath{\,M_{\rm J}}}
\providecommand{\rj}{\ensuremath{\,R_{\rm J}}}

\providecommand{\fave}{\langle F \rangle}
\providecommand{\fluxcgs}{10$^9$ erg s$^{-1}$ cm$^{-2}$}

\shorttitle{TKS. VIII. A TESS Giant Planet Orbiting at {0.83} au}
\shortauthors{Dalba et al.}

\begin{document}

\title{The TESS-Keck Survey. VIII. Confirmation of a Transiting Giant Planet on an Eccentric {261} day Orbit with the Automated Planet Finder Telescope \footnote{Some of the data presented herein were obtained at the W. M. Keck Observatory, which is operated as a scientific partnership among the California Institute of Technology, the University of California and the National Aeronautics and Space Administration. The Observatory was made possible by the generous financial support of the W. M. Keck Foundation.}}

\correspondingauthor{Paul A. Dalba}
\email{pdalba@ucr.edu}


\author[0000-0002-4297-5506]{Paul A.\ Dalba} 
\altaffiliation{NSF Astronomy and Astrophysics Postdoctoral Fellow}
\affiliation{Department of Astronomy and Astrophysics, University of California, Santa Cruz, CA 95064, USA}
\affiliation{Department of Earth and Planetary Sciences, University of California Riverside, 900 University Ave, Riverside, CA 92521, USA}

\author[0000-0002-7084-0529]{Stephen R.\ Kane} 
\affiliation{Department of Earth and Planetary Sciences, University of California Riverside, 900 University Ave, Riverside, CA 92521, USA}

\author[0000-0003-2313-467X]{Diana Dragomir} 
\affiliation{Department of Physics \& Astronomy, University of New Mexico, 1919 Lomas Blvd NE, Albuquerque, NM 87131, USA}

\author[0000-0001-6213-8804]{Steven Villanueva Jr.} 
\altaffiliation{Pappalardo Fellow}
\affiliation{Department of Physics and Kavli Institute for Astrophysics and Space Research, Massachusetts Institute of Technology, Cambridge, MA 02139, USA}

\author[0000-0001-6588-9574]{Karen A.\ Collins} 
\affiliation{Center for Astrophysics \textbar \ Harvard \& Smithsonian, 60 Garden St, Cambridge, MA 02138, USA}

\author[0000-0003-3988-3245]{Thomas Lee Jacobs} 
\altaffiliation{Citizen Scientist}
\affiliation{12812 SE 69th Place, Bellevue, WA 98006, USA}

\author[0000-0002-8527-2114]{Daryll M.\ LaCourse} 
\altaffiliation{Citizen Scientist}
\affiliation{7507 52nd Pl NE, Marysville, WA 98270, USA}

\author[0000-0002-5665-1879]{Robert Gagliano} 
\altaffiliation{Citizen Scientist}
\affiliation{Glendale, AZ 85308, USA}

\author[0000-0002-2607-138X]{Martti H.\ Kristiansen} 
\affiliation{Brorfelde Observatory, Observator Gyldenkernes Vej 7, DK-4340 T\o{}ll\o{}se, Denmark}
\affiliation{DTU Space, National Space Institute, Technical University of Denmark, Elektrovej 327, DK-2800 Lyngby, Denmark}

\author{Mark Omohundro} 
\altaffiliation{Citizen Scientist}
\affiliation{Department of Physics, University of Oxford, Denys Wilkinson Building, Keble Road, Oxford, OX13RH, UK}

\author{Hans M.\ Schwengeler} 
\altaffiliation{Citizen Scientist}
\affiliation{Planet Hunter, Bottmingen, Switzerland}

\author[0000-0002-0654-4442]{Ivan A.\ Terentev} 
\altaffiliation{Citizen Scientist}
\affiliation{Planet Hunter, Petrozavodsk, Russia}

\author[0000-0001-7246-5438]{Andrew Vanderburg} 
\affiliation{Department of Physics and Kavli Institute for Astrophysics and Space Research, Massachusetts Institute of Technology, Cambridge, MA 02139, USA}

\author[0000-0003-3504-5316]{Benjamin Fulton} 
\affiliation{NASA Exoplanet Science Institute/Caltech-IPAC, MC 314-6, 1200 E. California Blvd., Pasadena, CA 91125, USA}

\author[0000-0002-0531-1073]{Howard Isaacson} 
\affiliation{Department of Astronomy, University of California Berkeley, Berkeley CA 94720, USA}
\affiliation{Centre for Astrophysics, University of Southern Queensland, Toowoomba, QLD, Australia}

\author[0000-0002-4290-6826]{Judah Van Zandt} 
\affiliation{Department of Physics \& Astronomy, University of California Los Angeles, Los Angeles, CA 90095, USA}

\author[0000-0001-8638-0320]{Andrew W. Howard} 
\affiliation{Department of Astronomy, California Institute of Technology, Pasadena, CA 91125, USA}

\author[0000-0002-5113-8558]{Daniel P.\ Thorngren} 
\affiliation{Institute for Research on Exoplanets (iREx), Universit\'e de Montr\'eal, Canada}

\author[0000-0002-2532-2853]{Steve B. Howell} 
\affiliation{NASA Ames Research Center, Moffett Field, CA 94035, USA}


\author[0000-0002-7030-9519]{Natalie M. Batalha} 
\affiliation{Department of Astronomy and Astrophysics, University of California, Santa Cruz, CA 95064, USA}

\author[0000-0003-1125-2564]{Ashley Chontos} 
\altaffiliation{NSF Graduate Research Fellow}
\affiliation{Institute for Astronomy, University of Hawai`i, 2680 Woodlawn Drive, Honolulu, HI 96822, USA}

\author{Ian J. M. Crossfield} 
\affiliation{Department of Physics \& Astronomy, University of Kansas, 1082 Malott, 1251 Wescoe Hall Dr., Lawrence, KS 66045, USA}

\author[0000-0001-8189-0233]{Courtney D.\ Dressing} 
\affiliation{Department of Astronomy, University of California Berkeley, Berkeley CA 94720, USA}

\author[0000-0001-8832-4488]{Daniel Huber} 
\affiliation{Institute for Astronomy, University of Hawai`i, 2680 Woodlawn Drive, Honolulu, HI 96822, USA}

\author[0000-0003-0967-2893]{Erik A.\ Petigura} 
\affiliation{Department of Physics \& Astronomy, University of California Los Angeles, Los Angeles, CA 90095, USA}

\author[0000-0003-0149-9678]{Paul Robertson} 
\affiliation{Department of Physics \& Astronomy, University of California Irvine, Irvine, CA 92697, USA}

\author[0000-0001-8127-5775]{Arpita Roy} 
\affiliation{Space Telescope Science Institute, 3700 San Martin Drive, Baltimore, MD 21218, USA}
\affiliation{Department of Physics and Astronomy, Johns Hopkins University, 3400 N Charles St, Baltimore, MD 21218, USA}

\author[0000-0002-3725-3058]{Lauren M.\ Weiss} 
\affiliation{Department of Physics, University of Notre Dame, Notre Dame, IN 46556, USA}


\author[0000-0003-0012-9093]{Aida Behmard} 
\altaffiliation{NSF Graduate Research Fellow}
\affiliation{Division of Geological and Planetary Science, California Institute of Technology, Pasadena, CA 91125, USA}

\author[0000-0001-7708-2364]{Corey Beard} 
\affiliation{Department of Physics \& Astronomy, University of California Irvine, Irvine, CA 92697, USA}

\author[0000-0002-4480-310X]{Casey L.\ Brinkman} 
\affiliation{Institute for Astronomy, University of Hawai`i, 2680 Woodlawn Drive, Honolulu, HI 96822, USA}

\author[0000-0002-8965-3969]{Steven Giacalone} 
\affiliation{Department of Astronomy, University of California Berkeley, Berkeley CA 94720, USA}

\author[0000-0002-0139-4756]{Michelle L.\ Hill} 
\affiliation{Department of Earth and Planetary Sciences, University of California Riverside, 900 University Ave, Riverside, CA 92521, USA}

\author[0000-0001-8342-7736]{Jack Lubin} 
\affiliation{Department of Physics \& Astronomy, University of California Irvine, Irvine, CA 92697, USA}

\author[0000-0002-7216-2135]{Andrew W.\ Mayo} 
\affiliation{Department of Astronomy, University of California Berkeley, Berkeley CA 94720, USA}
\affiliation{Centre for Star \& Planet Formation, Natural History Museum of Denmark \& Niels Bohr Institute, University of Copenhagen, \O{}ster Voldgade 5-7, DK-1350 Copenhagen K, Denmark}

\author[0000-0003-4603-556X]{Teo Mo\v{c}nik} 
\affiliation{Gemini Observatory/NSF's NOIRLab, 670 N. A'ohoku Place, Hilo, HI 96720, USA}

\author[0000-0001-8898-8284]{Joseph M.\ Akana Murphy} 
\altaffiliation{NSF Graduate Research Fellow, LSSTC Data Science Fellow}
\affiliation{Department of Astronomy and Astrophysics, University of California, Santa Cruz, CA 95064, USA}

\author[0000-0001-7047-8681]{Alex S.\ Polanski} 
\affiliation{Department of Physics \& Astronomy, University of Kansas, 1082 Malott, 1251 Wescoe Hall Dr., Lawrence, KS 66045, USA}

\author[0000-0002-7670-670X]{Malena Rice} 
\altaffiliation{NSF Graduate Research Fellow}
\affiliation{Department of Astronomy, Yale University, New Haven, CT 06511, USA}

\author[0000-0001-8391-5182]{Lee J.\ Rosenthal} 
\affiliation{Department of Astronomy, California Institute of Technology, Pasadena, CA 91125, USA}

\author[0000-0003-3856-3143]{Ryan A.\ Rubenzahl} 
\altaffiliation{NSF Graduate Research Fellow}
\affiliation{Department of Astronomy, California Institute of Technology, Pasadena, CA 91125, USA}

\author[0000-0003-3623-7280]{Nicholas Scarsdale}  
\affiliation{Department of Astronomy and Astrophysics, University of California, Santa Cruz, CA 95064, USA}

\author{Emma V.\ Turtelboom} 
\affiliation{Department of Astronomy, University of California Berkeley, Berkeley CA 94720, USA}

\author{Dakotah Tyler} 
\affiliation{Department of Physics \& Astronomy, University of California Los Angeles, Los Angeles, CA 90095, USA}


\author[0000-0001-6981-8722]{Paul Benni} 
\altaffiliation{Citizen Scientist}
\affiliation{Acton Sky Portal (Private Observatory), Acton, MA, USA}

\author{Pat Boyce} 
\affiliation{Boyce Research Initiatives and Education Foundation, 3540 Carleton Street, San Diego, CA, 92106}

\author[0000-0002-0792-3719]{Thomas M.\ Esposito} 
\affiliation{SETI Institute, Carl Sagan Center, 189 Bernardo Ave, Suite 200, Mountain View, CA 94043, USA}
\affiliation{Department of Astronomy, University of California Berkeley, Berkeley CA 94720, USA}

\author[0000-0002-5443-3640]{E. Girardin} 
\altaffiliation{Citizen Scientist}
\affiliation{Grand-Pra private Observatory, 1984 Les Hauderes, Switzerland}

\author{Didier Laloum} 
\altaffiliation{Citizen Scientist}
\affiliation{Observatoire Priv\'e du Mont, 40280 Saint-Pierre-du-Mont, France}

\author[0000-0003-0828-6368]{Pablo Lewin} 
\altaffiliation{Citizen Scientist}
\affiliation{The Maury Lewin Astronomical Observatory, Glendora, CA 91741, USA}

\author[0000-0002-9312-0073]{Christopher R.\ Mann} 
\affiliation{Institute for Research on Exoplanets (iREx), Universit\'e de Montr\'eal, Canada}

\author[0000-0001-7016-7277]{Franck Marchis} 
\affiliation{SETI Institute, Carl Sagan Center, 189 Bernardo Avenue, Mountain View, CA, USA}
\affiliation{Unistellar, 198 Alabama Street, San Francisco CA 94110 USA}

\author[0000-0001-8227-1020]{Richard P.\ Schwarz} 
\affiliation{Patashnick Voorheesville Observatory, Voorheesville, NY 12186, USA}

\author{Gregor Srdoc} 
\altaffiliation{Citizen Scientist} 
\affiliation{Kotizarovci Observatory, Sarsoni 90, 51216 Viskovo, Croatia}

\author{Jana Steuer} 
\affiliation{University Observatory Munich (USM), Scheinerstra\ss e 1, D-81679 Munich, Germany}

\author{Thirupathi Sivarani} 
\affiliation{Indian Institute of Astrophysics, 2nd block Koramangala, Bangalore-34, India}

\author{Athira Unni} 
\affiliation{Indian Institute of Astrophysics, 2nd block Koramangala, Bangalore-34, India}


\author[0000-0002-9138-9028]{Nora L.\ Eisner} 
\affiliation{Department of Physics, University of Oxford, Keble Road, Oxford OX3 9UU, UK}

\author[0000-0002-3551-279X]{Tara Fetherolf} 
\altaffiliation{UC Chancellor's Fellow}
\affiliation{Department of Earth and Planetary Sciences, University of California Riverside, 900 University Ave, Riverside, CA 92521, USA}

\author[0000-0002-4860-7667]{Zhexing Li} 
\affiliation{Department of Earth and Planetary Sciences, University of California Riverside, 900 University Ave, Riverside, CA 92521, USA}

\author[0000-0003-4554-5592]{Xinyu Yao}  
\affiliation{Department of Physics, Lehigh University, 16 Memorial Drive East, Bethlehem, PA 18015, USA}
\affiliation{Shanghai Astronomical Observatory, Chinese Academy of Sciences, 80 Nandan Road, Shanghai 200030, China}

\author[0000-0002-3827-8417]{Joshua Pepper} 
\affiliation{Department of Physics, Lehigh University, 16 Memorial Drive East, Bethlehem, PA 18015, USA}


\author[0000-0003-2058-6662]{George R.\ Ricker} 
\affiliation{Department of Physics and Kavli Institute for Astrophysics and Space Research, Massachusetts Institute of Technology, Cambridge, MA 02139, USA}

\author[0000-0001-6763-6562]{Roland Vanderspek} 
\affiliation{Department of Physics and Kavli Institute for Astrophysics and Space Research, Massachusetts Institute of Technology, Cambridge, MA 02139, USA}

\author[0000-0001-9911-7388]{David W.\ Latham} 
\affiliation{Center for Astrophysics \textbar \ Harvard \& Smithsonian, 60 Garden St, Cambridge, MA 02138, USA}

\author[0000-0002-6892-6948]{S. Seager} 
\affiliation{Department of Physics and Kavli Institute for Astrophysics and Space Research, Massachusetts Institute of Technology, Cambridge, MA 02139, USA}
\affiliation{Department of Earth, Atmospheric and Planetary Sciences, Massachusetts Institute of Technology, Cambridge, MA 02139, USA}
\affiliation{Department of Aeronautics and Astronautics, MIT, 77 Massachusetts Avenue, Cambridge, MA 02139, USA}

\author[0000-0002-4265-047X]{Joshua N.\ Winn} 
\affiliation{Department of Astrophysical Sciences, Princeton University, 4 Ivy Lane, Princeton, NJ 08544, USA}

\author[0000-0002-4715-9460]{Jon M.\ Jenkins} 
\affiliation{NASA Ames Research Center, Moffett Field, CA 94035, USA}


\author[0000-0002-7754-9486]{Christopher J.\ Burke} 
\affiliation{Department of Physics and Kavli Institute for Astrophysics and Space Research, Massachusetts Institute of Technology, Cambridge, MA 02139, USA}

\author[0000-0003-3773-5142]{Jason D.\ Eastman} 
\affiliation{Center for Astrophysics \textbar \ Harvard \& Smithsonian, 60 Garden St, Cambridge, MA 02138, USA}

\author[0000-0003-2527-1598]{Michael B.\ Lund} 
\affiliation{Caltech IPAC – NASA Exoplanet Science Institute, 1200 E. California Ave, Pasadena, CA 91125, USA}

\author{David R.\ Rodriguez} 
\affiliation{Space Telescope Science Institute, 3700 San Martin Drive, Baltimore, MD, 21218, USA}

\author[0000-0002-4829-7101]{Pamela Rowden} 
\affiliation{Royal Astronomical Society, Burlington House, Piccadilly, London W1J 0BQ, UK}

\author[0000-0002-8219-9505]{Eric B.\ Ting} 
\affiliation{NASA Ames Research Center, Moffett Field, CA 94035, USA}

\author{Jesus Noel Villase{\~n}or} 
\affiliation{Department of Physics and Kavli Institute for Astrophysics and Space Research, Massachusetts Institute of Technology, Cambridge, MA 02139, USA}


\begin{abstract}
We report the discovery of TOI-2180 b, a {2.8 $M_{\rm J}$} giant planet orbiting a slightly evolved G5 host star. This planet transited only once in Cycle 2 of the primary Transiting Exoplanet Survey Satellite (TESS) mission. Citizen scientists identified the 24 hr single-transit event shortly after the data were released, allowing a Doppler monitoring campaign with the Automated Planet Finder telescope at Lick Observatory to begin promptly. The radial velocity observations refined the orbital period of TOI-2180 b to be {260.8$\pm$0.6} days, revealed an orbital eccentricity of {0.368$\pm$0.007}, and discovered long-term acceleration from a more distant massive companion. We conducted ground-based photometry from 14 sites spread around the globe in an attempt to detect another transit. Although we did not make a clear transit detection, the nondetections improved the precision of the orbital period. We predict that TESS will likely detect another transit of TOI-2180 b in Sector 48 of its extended mission. We use giant planet structure models to retrieve the bulk heavy-element content of TOI-2180 b. When considered alongside other giant planets with orbital periods over 100 days, we find tentative evidence that the correlation between planet mass and metal enrichment relative to stellar is dependent on orbital properties. Single-transit discoveries like TOI-2180 b highlight the exciting potential of the TESS mission to find planets with long orbital periods and low irradiation fluxes despite the selection biases associated with the transit method. 
\end{abstract}

 
\section{Introduction}\label{sec:intro}

Gas giant planets have been found to reside in many extrasolar planetary systems. The diversity in their sizes, masses, orbits, compositions, and formation pathways has been the subject of numerous studies. However, selection biases often cloud our understanding. For instance, the sensitivity of the transit method wanes for planets on orbits beyond a few tenths of an au owing to the inverse relation between transit probability and semi-major axis. Consequently, the vast majority of known exoplanets with $\gtrsim$1~au orbits have been discovered via Doppler spectroscopy \citep[e.g.,][]{Mayor2011,Fulton2021}. Planet mass can be inferred from time series radial velocity (RV) observations but, without a transit, the planet radius and thereby bulk density remains unknown. 

A critical component of the bulk composition for giant planets is the total mass of elements heavier than H and He \citep[e.g.,][]{Guillot2006,Miller2011,Thorngren2016,Teske2019}. This property is inferred using structural evolution models along with the measured planetary mass, radius, stellar age, and incident flux \citep[e.g.,][]{Thorngren2019a}. Numerous theoretical planet formation studies have found that the correlation between the total mass of heavy elements in giant planets---or, similarly, the metal enrichment relative to the host star---and planet mass is a useful tracer of planet formation processes \citep[e.g.,][]{Mordasini2014,Hasegawa2018,Ginzburg2020,Shibata2020}. Probing this giant planet mass--metallicity correlation along a third axis, in orbital properties (i.e., period or separation), could prove informative. Yet, previous efforts have simply not had a large enough sample size of giant planets on orbits wider than a few 0.1~au to conduct such an investigation \citep{Miller2011,Thorngren2016}.

Transit surveys such as the \kepler\ mission \citep{Borucki2010,Thompson2018} and the Transiting Exoplanet Survey Satellite (TESS) mission \citep{Ricker2015} occasionally detect giant planet candidates with orbital periods of 100 to 1,000 days \citep[e.g.,][]{Wang2015b,ForemanMackey2016b,Osborn2016,Uehara2016,Herman2019,Kawahara2019,Eisner2021}. These candidates warrant close scrutiny, at least based on the $\sim$50\% false-positive rate for giant planets with similar orbits found for the \kepler\ mission \citep{Santerne2014,Dalba2020b}. Following vetting, long-term RV monitoring is often required to measure the planet mass, which then enables modeling of the bulk metallicity \citep[e.g.,][]{Beichman2016,Dubber2019,Santerne2019,Dalba2021a,Dalba2021c}. \tess\ planets with orbital periods on the order of 100~days or more will typically be detected as single-transit events owing to the observational strategy of the TESS mission \citep[e.g.,][]{LaCourse2018,Villanueva2019,Diaz2020,Cooke2021}. In this case, the RV monitoring is also necessary to determine the orbital period. 

Here, we describe the discovery and investigation of a 24-hour single-transit event observed for HD~238894, hereafter referred to as \planet. The host star, \host\ (TIC~298663873), is a bright ($V=9.2$), slightly evolved G5 star. The confirmation of \planet\ is notable relative to the current sample of TESS exoplanets, including those identified as single-transits, as it is the first to surpass an orbital period of 100~days\footnote{According to the list of confirmed TESS exoplanets on the NASA Exoplanet Archive (\url{https://exoplanetarchive.ipac.caltech.edu/index.html}) as of 2021 September 2.}. In Section~\ref{sec:obs}, we describe the \tess\ observations of \host\ along with the follow-up photometric, imaging, and spectroscopic data sets. In Section~\ref{sec:model}, we discuss the consistency of the transit and RV data, which are then jointly modeled to determine the \host\ system properties. We find that \planet\ is a {2.7~$M_{\rm J}$} giant planet on a {261~day} orbit with an orbital eccentricity of {0.368}. In Section~\ref{sec:teers}, we describe a global effort to detect an additional transit of \planet\ from the ground. Although this effort failed to detect the transit, the nondetections provided a substantial improvement in the precision on the orbital period. In Section~\ref{sec:metal}, we determine the bulk heavy-element mass of \planet. In Section~\ref{sec:drift}, we use the long-term acceleration of \host\ to place limits on the properties of a distant massive companion. In Section~\ref{sec:disc}, we place \planet\ in the context of other transiting giant planets and discuss the prospects for future characterization of the system. Lastly, in Section~\ref{sec:concl}, we summarize our work.


\section{Observations}\label{sec:obs}

\subsection{TESS Photometry}\label{sec:tess}

TESS observed \host\ (TIC~298663873) at 2-min (fast) cadence for the entirety of Cycle~2 of its primary mission (Sectors~14--26) and in Sectors~40 and 41 of its extended mission. The image data were processed by the Science Processing Operations Center (SPOC) at NASA Ames Research Center \citep{Jenkins2016} to extract photometry from this target. A search for transiting planets failed to find a transit signature with two or more transits. Only a single transit event was observed in Sector~19 (2019 November 28 through 2019 December 22). This Sector~19 single-transit was first identified by citizen scientists with the light curve processing software \texttt{LcTools} \citep{Schmitt2019}, leading to early commencement of our RV follow-up campaign. In May 2020, the Planet Hunters TESS collaboration \citep{Eisner2020} announced this star as a Community TESS Object of Interest (cTOI). In August 2020, its disposition was elevated to TOI \citep{Guerrero2021}.

\begin{figure}
    \centering
    \begin{tabular}{c}
    \includegraphics[width=\columnwidth]{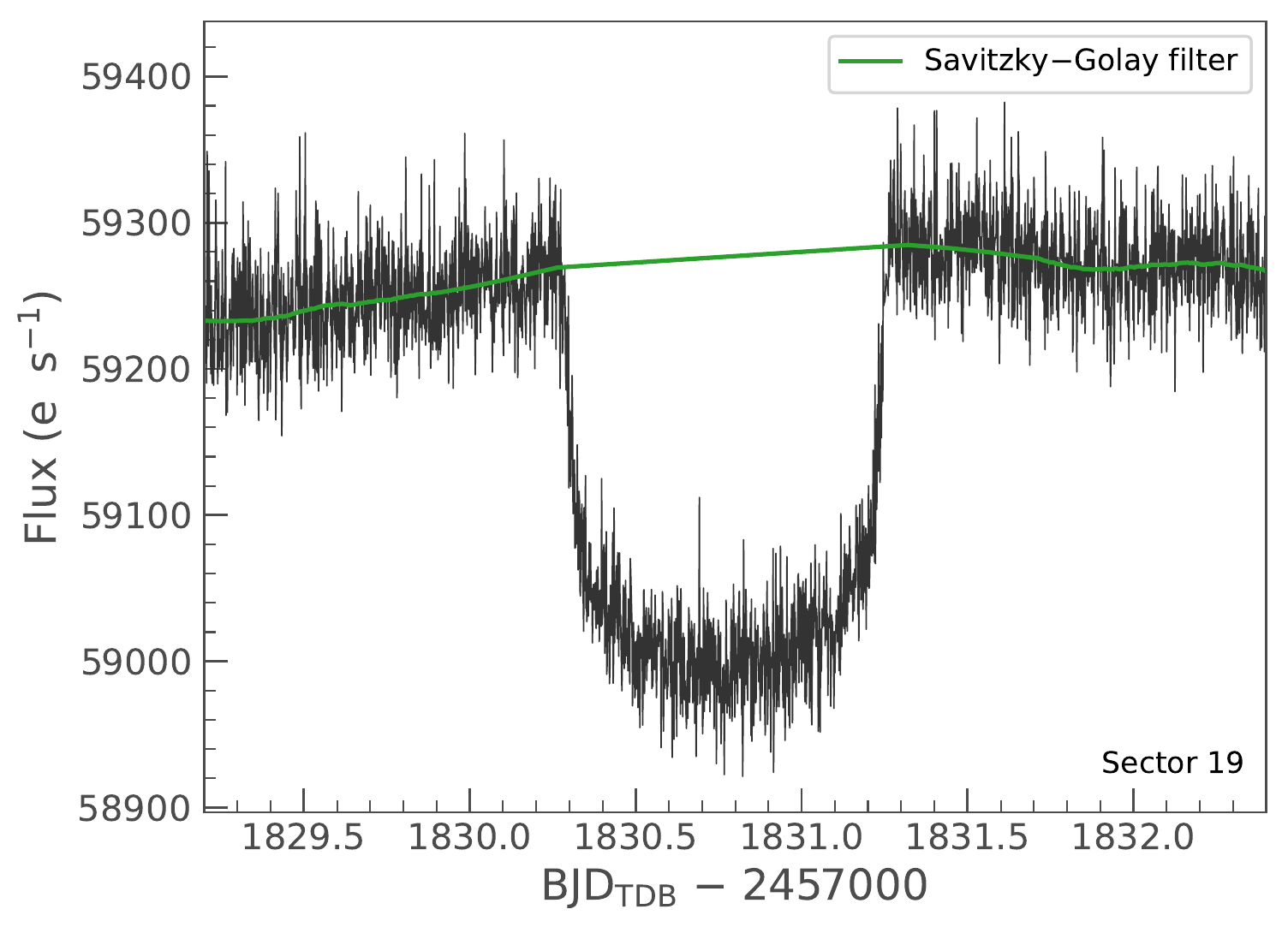} \\
    \includegraphics[width=\columnwidth]{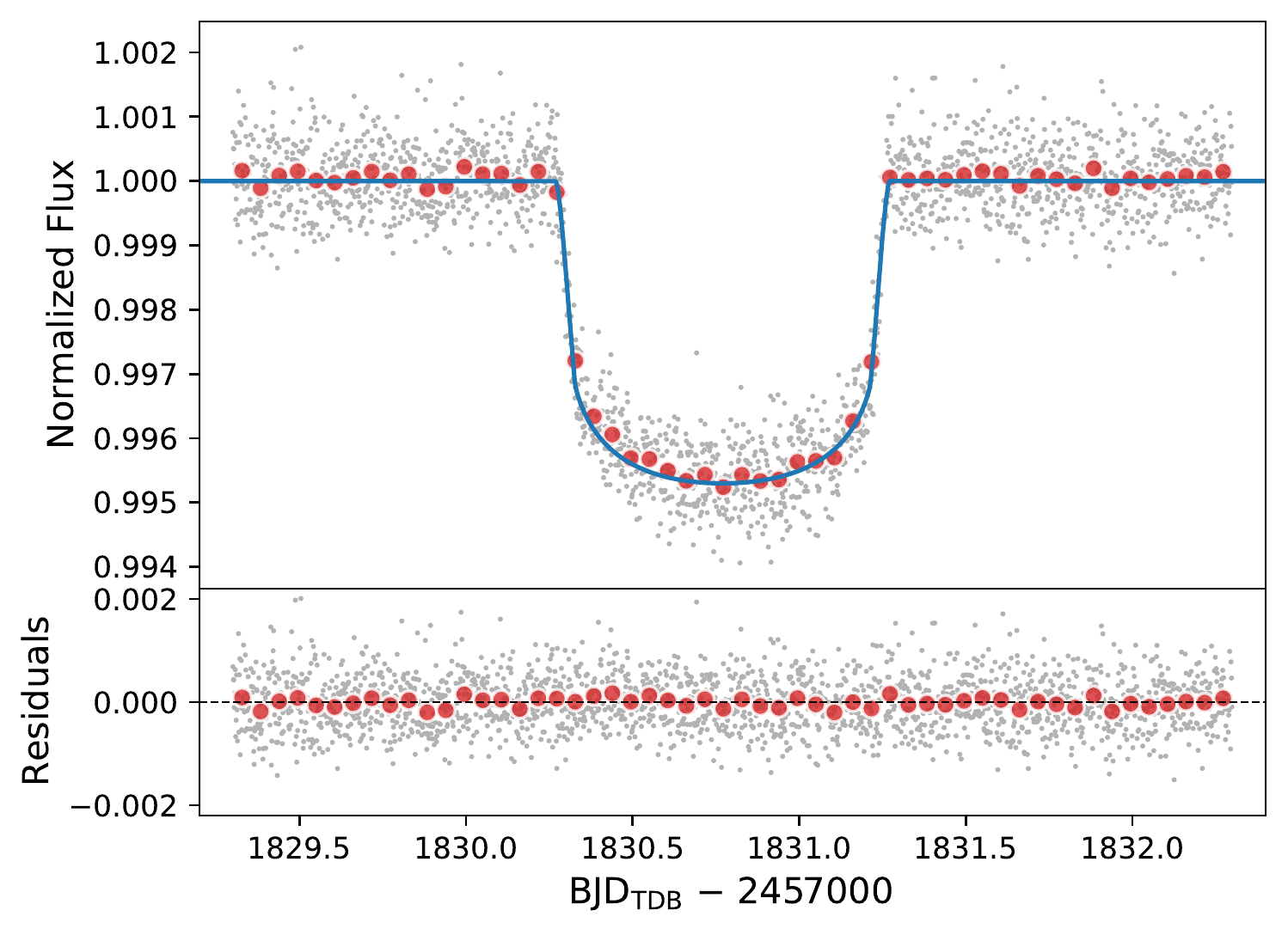}
    \end{tabular}
    \caption{The single-transit of \planet\ observed by TESS with short cadence. {\it Top:} unflattened PDCSAP flux and the trend from the Savitsky--Golay filter. {\it Bottom:} flattened light curve. The red points are individual exposures (gray points) binned by a factor of 40. The blue line shows the best fit transit model.}
    \label{fig:tess}
\end{figure}

We downloaded the Sector~19, 2-min cadence light curve of \host\ from the Mikulski Archive for Space Telescopes (MAST) using the \texttt{lightkurve} package \citep{Lightkurve2018}. We used the pre-search data conditioning simple aperture photometry (PDCSAP) flux for which most of the high frequency noise was removed \citep{Stumpe2012,Stumpe2014,Smith2012}. The PDCSAP light curve exhibited low frequency noise features that we removed using a Savitzky--Golay filter after masking the clear transit event. The raw and flattened light curves, centered on the $\sim$0.5\% transit of \planet, are shown in Figure~\ref{fig:tess}. 

Under the assumption of a circular central transit,  \citet[][Equation~19]{Winn2010b} showed how the transit duration ($T$) and the stellar bulk density ($\rho_{\star}$) can give an estimate of the orbital period ($P$) following
\begin{equation}\label{eq:P}
T \approx 13\;\mathrm{hr} \; \left (\frac{P}{1\; \mathrm{yr}}   \right )^{1/3} \left (\frac{\rho_{\star}}{\rho_{\sun}}  \right )^{-1/3}
\end{equation}

\noindent where $P$ has units of years and $\rho_{\sun}$ is the solar bulk density. For $T=24$~hr and $\rho_{\star}=0.335$~g~cm$^{-3}$ from the TESS Input Catalog \citep[TIC;][]{Stassun2019}, Equation~\ref{eq:P} gives $P\approx547\pm111$~days assuming reasonable uncertainty in $T$ and $\rho_{\star}$. Combined with the nondetection of a matching transit event in the other Cycle 2 sectors, this possible orbital period suggested that \planet\ was likely to be one of only a few long-period planets predicted to be detected by TESS through single-transit events \citep{Villanueva2019}. 

\subsection{Speckle Imaging}\label{sec:imaging}

\begin{figure}
    \centering
    \includegraphics[width=\columnwidth]{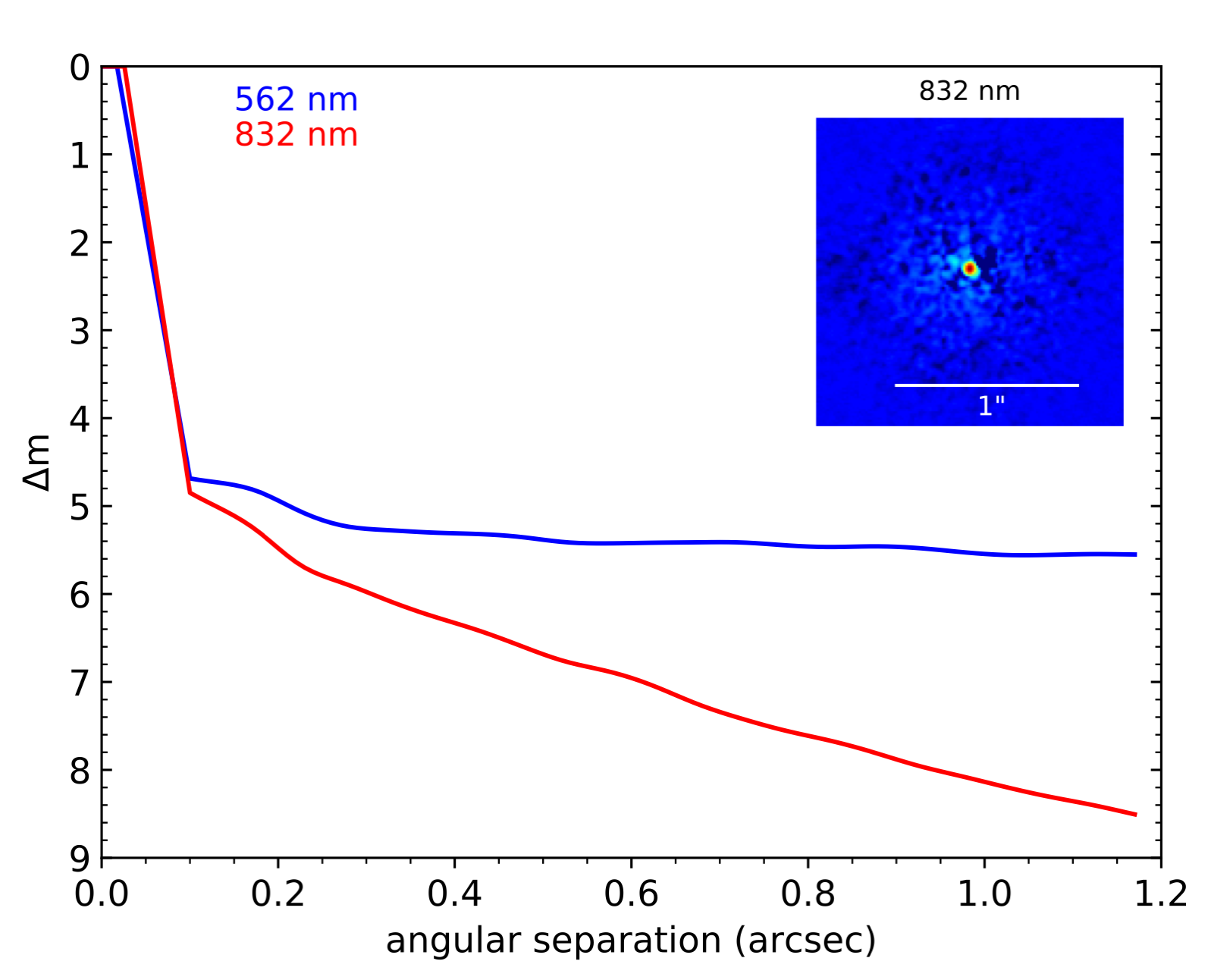}
    \caption{Limiting magnitudes ($\Delta m$) for the nondetection of a neighboring star based on the speckle imaging from `Alopeke. The inset shows the speckle image at 832~nm.}
    \label{fig:speckle}
\end{figure}

We acquired a high-resolution speckle image of \host\ to search for nearby neighboring stars that might indicate the false-positive nature of the TESS single-transit event. We observed \host\ on 2020 June 6 using the `Alopeke
speckle instrument on the Gemini-North telescope\footnote{\url{https://www.gemini.edu/instrumentation/alopeke-zorro}} located on Maunakea in Hawai`i. `Alopeke acquires an image of the star in a blue (562~nm) and red (832~nm) band simultaneously. From these images, we derived contrast curves that show the limiting magnitude difference ($\Delta m$) in each band as a function of angular separation \citep{Howell2011}. As shown in Figure~\ref{fig:speckle}, we achieved a $\sim$5~mag contrast at 0$\farcs$1 and a nondetection of sources with $\Delta m$ = 5--8.5 within 1$\farcs$2 of \host. Given the distance to \host\ ({116~pc}; Table~\ref{tab:stellar}), 0$\farcs$1 projected separation corresponds to {$\sim$12~au}. Although we cannot rule out a scenario whereby a luminous companion star was near conjunction with \host\ at the time of our speckle observation, we assume in what follows that \host\ is likely a single star. The speckle imaging nondetection suggests that any massive object in this system other than \planet\ is likely to be a late M star or a substellar object. 

The conclusion of \host\ being a single star is further supported by the renormalized unit weight error (RUWE) as determined by Gaia Data Release (DR) 2 \citep{Gaia2018}. The RUWE value for \host\ is 1.01, which is typical of a single star \citep[e.g.,][]{Belokurov2020}.

\subsection{Spectroscopy}\label{sec:spec}

Immediately following the discovery of the TESS single-transit of \host, we began a Doppler monitoring campaign with the Automated Planet Finder (APF) telescope at Lick Observatory \citep{Radovan2014,Vogt2014} as part of the TESS-Keck Survey (TKS). TKS is a collaborative effort between the University of California, the California Institute of Technology, the University of Hawai`i,
and NASA aimed at providing multi-site RV follow-up for many of the planetary systems discovered by TESS \citep[e.g.,][]{Dai2020,Dalba2020a,Chontos2021,Lubin2021,Rubenzahl2021,Weiss2021}. The APF uses the Levy Spectrograph, a high-resolution ($R\approx$ 114,000) slit-fed optical echelle spectrometer \citep{Radovan2010} that is ideal for bright (V$\le$9) stars such as \host\ \citep{Burt2015}. The starlight passes through a heated iodine gas cell that allows for precise wavelength calibration and instrument profile tracking. The precise RV is inferred from each spectrum using a forward modeling procedure \citep{Butler1996,Fulton2015a}. Table~\ref{tab:rvs} lists the RVs collected in our campaign.

\begin{deluxetable}{cccc}
\tabletypesize{\scriptsize}
\tablecaption{RV Measurements of \host \label{tab:rvs}}
\tablehead{
  \colhead{BJD$_{\rm TDB}$} & 
  \colhead{RV (m s$^{-1}$)} &
  \colhead{$S_{\rm HK}$$^a$} &
  \colhead{Tel.}}
  \startdata
    2458888.063868 & $-16.1\pm3.9$ & $0.1304\pm0.0020$ & APF\\
    2458894.911472 & $-29.1\pm4.5$ & $0.1476\pm0.0020$ & APF\\
    2458899.027868 & $-36.3\pm3.8$ & $0.1463\pm0.0020$ & APF\\
    2458906.015644 & $-43.6\pm3.2$ & $0.1093\pm0.0020$ & APF\\
    2458914.011097 & $-38.9\pm3.3$ & $0.1506\pm0.0020$ & APF\\
    2458954.765221 & $-59.0\pm4.0$ & $0.1468\pm0.0020$ & APF\\
    2458961.864505 & $-49.7\pm4.3$ & $0.1323\pm0.0020$ & APF\\
    2458964.879814 & $-53.3\pm3.1$ & $0.1298\pm0.0020$ & APF\\
    2458965.811833 & $-47.5\pm4.2$ & $0.1363\pm0.0020$ & APF\\
    2458966.879965 & $-65.9\pm5.5$ & $0.1204\pm0.0020$ & APF\\
\enddata
\tablenotetext{a}{The $S_{\rm HK}$ values from APF and Keck data have different zero-points.}
\tablenotetext{}{This is a representative subset of the full data set. The full table will be made available in machine readable format.}
\end{deluxetable}

In August 2020, Lick Observatory and the APF shut down owing to nearby wildfires. To maintain our coverage of the emerging Keplerian RV signal, we temporarily conducted observations of \host\ using the High Resolution Echelle Spectrometer \citep[HIRES;][]{Vogt1994} on the Keck~I telescope at W. M. Keck Observatory. The reduction and analysis procedure for Keck-HIRES spectra is broadly similar to that for data from the APF \citep[e.g.,][]{Howard2010}. The full version of Table~\ref{tab:rvs} also contains the Keck-HIRES RVs of \host.   

\begin{figure*}
    \centering
    \includegraphics[width=\textwidth]{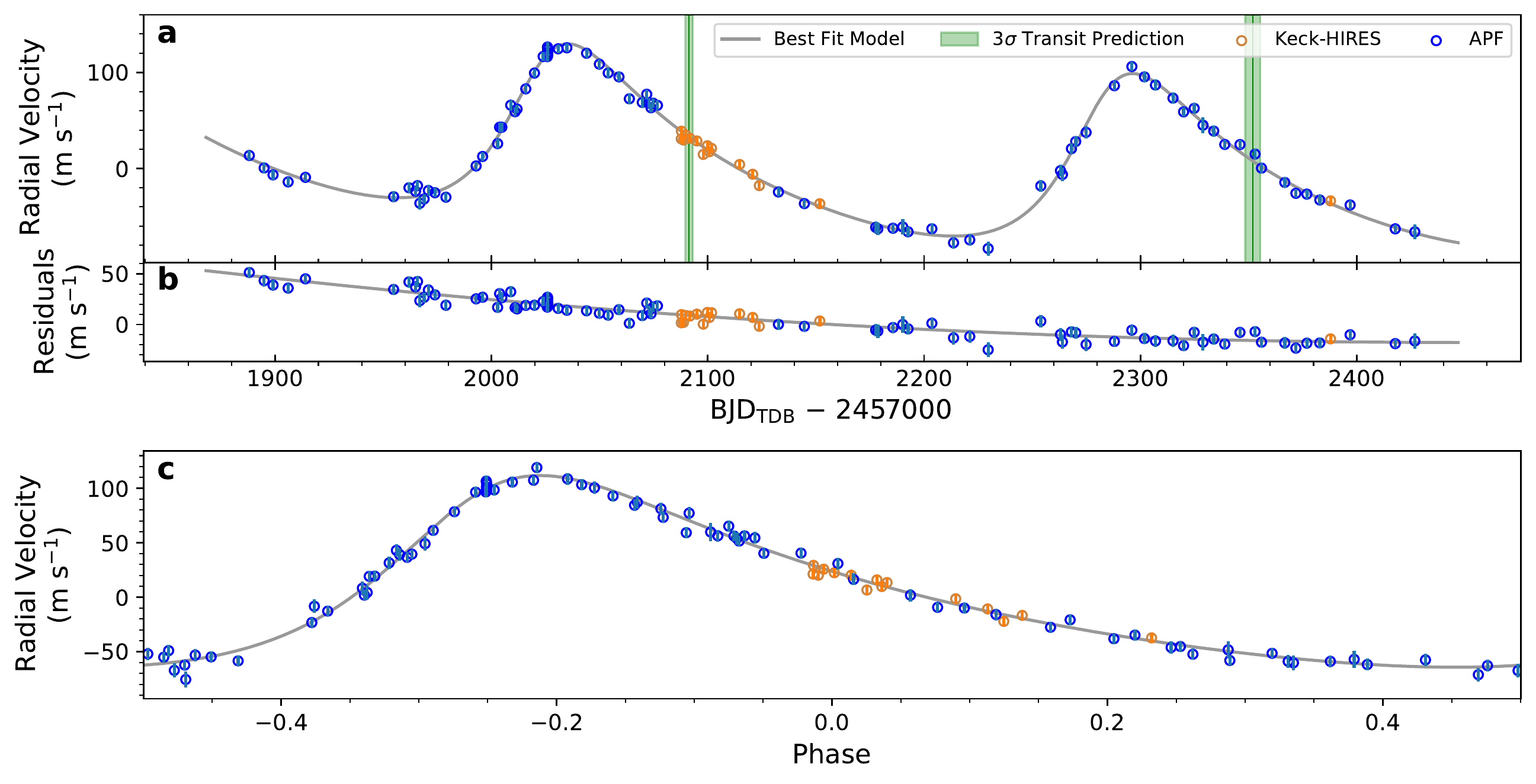}
    \caption{RV measurements of \host. {\it Panel a:} the RV time series after subtraction of an offset velocity from each data set. Transit windows are shown in green. {\it Panel b:} the residuals between the RV time series and best fit planet model not including acceleration terms. A quadratic trend is visible on top of the Keplerian signal from \planet. {\it Panel c:} the phase-folded RVs after removal of the acceleration terms. A phase of 0 corresponds to conjunction (transit).}
    \label{fig:rv}
\end{figure*}

The time series RVs from both telescopes are shown in the top panel of Figure~\ref{fig:rv}. Our {1.5~yr} baseline captured {two} periods of a 261~day, eccentric ($e\approx0.4$) Keplerian signal as well as a long-term acceleration. In Section~\ref{sec:model} we will discuss the consistency of the RV data with the single-transit from TESS and conduct a joint modeling of the system parameters. 

Each RV measurement is accompanied by an estimate of stellar chromospheric activity approximated as the $S_{\rm HK}$ index \citep{Isaacson2010}. The indicators calculated for each instrument have different zero points. The Pearson correlation coefficient between the $S_{\rm HK}$ indices and the RVs for APF and Keck are {$-0.10$ (from 84 data points)} and {$0.68$ (from 15 data points)}, respectively. Although the Keck data demonstrate a positive correlation significant to {3.3$\sigma$}, the much larger APF data set shows no correlation {($<1\sigma$)}. Furthermore, we do not detect significant periodicity in the $S_{\rm HK}$ time series. Both of these findings suggest that the RV signals are not affected, or only minimally so, by stellar activity.  

The extraction of the APF and Keck RVs rely on a high signal-to-noise template spectrum of \host\ acquired with Keck-HIRES without the iodine cell \citep{Butler1996}. We also used this spectrum for a basic spectroscopic analysis of \host\ with \texttt{SpecMatch} \citep{Petigura2015,Petigura2017b}. This analysis indicated that \host\ has an effective temperature of $T_{\rm eff} = 5739\pm100$~K, a surface gravity of $\log{g} = 4.00\pm0.10$, and a relative iron abundance (metallicity proxy) of [Fe/H] = 0.25$\pm$0.06~dex consistent with a slightly evolved mid-G type star. These values are in close agreement with those listed in the TESS Input Catalog \citep{Stassun2019}.

\subsection{Ground-based Photometry}

The single-transit detection for \host\ by TESS and the subsequent RV follow-up effort allowed us to plan a ground-based photometry campaign with the goal of detecting another transit of \planet. This campaign occurred in late August and early September of 2020 around the time of the first transit of \planet\ since the one observed by TESS. We acquired 55 photometric data sets of \host\ comprised of $\sim$20,000 individual exposures spanning 11~days and 14 sites. Contributors to this campaign consisted of a mix of professional and amateur astronomers.  

The quality of the ground-based telescope data sets varied widely. Some data achieved precision sufficient to rule out ingress or egress events for the relatively shallow (0.5\%) transit of \planet. Other data showed correlated noise or systematic errors several times this magnitude. We leave a detailed discussion of the individual observing sites for Appendix~\ref{app:ground}. In Section~\ref{sec:teers}, we will discuss our treatment of the ground-based data as a whole and our procedure for using it to refine the transit ephemeris of \planet.


\section{Modeling}\label{sec:model}

\subsection{Consistency between the Transit and the RV Data}\label{sec:tran_rv}

A single-transit event only places a weak constraint on orbital period even under various simplifying assumptions \citep[e.g.,][]{Yee2008,Yao2021}. We must therefore assess whether the same object caused the single-transit and Doppler reflex motion of the host star.

In Section~\ref{sec:tess}, we estimated that the duration of the single-transit of \planet\ and the stellar density listed in the TIC corresponded to a 547-day orbital period assuming a circular, central transit. However, the RV measurements of \host\ suggest a shorter period that has significant eccentricity. Depending on the argument of periastron ($\omega_{\star}$), orbital eccentricity can lead to shorter or longer transit duration compared to that from a circular orbit \citep[e.g.,][]{Kane2012}, which can bias the orbital period estimation from a single-transit. Orbital eccentricity also increases the transit probability \citep{Kane2007}, which is inherently low for objects with au-scale orbital distances. 

We can account for eccentricity in our estimation of the orbital period to good approximation by including a factor of $\sqrt{1-e^2}/(1+e\sin{\omega_{\star}})$ on Equation~\ref{eq:P} \citep{Winn2010b}. We determined $e$ and $\omega_{\star}$ by conducting a Keplerian model fit to the APF and Keck RVs using the \texttt{RadVel} modeling toolkit\footnote{\url{https://radvel.readthedocs.io/}} \citep{Fulton2018}. In this fit, we applied a normal prior on the time of conjunction (BJD$_{\rm TDB} = 2458830.79\pm0.05$) using the transit timing from the Planet Hunters TESS characterization of the single-transit \citep{Eisner2021}. The fit converged quickly and we found that $e = 0.367\pm0.0074$, $\omega_{\star} = -0.76\pm0.023$~rad, and $P = 260.68 \pm 0.54$~days. This argument of periastron corresponds to a time of periastron that is $70.0\pm1.2$~days prior to transit. Therefore, at the time of transit, the orbital velocity of \planet\ is decreasing.  

Reevaluating Equation~\ref{eq:P} with the factor to account for an eccentric orbit gives $283\pm59$~days, which is consistent with the period of the Keplerian signal in the RVs. Considering the uncertainty introduced by the transit impact parameter could further explain the difference between this orbital period estimate and the observed {261}~day period. This result demonstrates self-consistency with our assumption of a T$_0$ value in the \texttt{RadVel} fit. Therefore, we continue our analysis with the implicit assumption that the single-transit in the photometry and the Keplerian signal in the RVs can be ascribed to the same planet: \planet.

\subsection{Comprehensive System Modeling}\label{sec:EFv2}

We modeled the stellar and planetary parameters for the \host\ system using the \texttt{EXOFASTv2} modeling suite \citep{Eastman2013,Eastman2019}. We include the TESS single-transit photometry, all of the RVs from Keck-HIRES and APF-Levy, and archival broadband photometry of \host\ from Gaia DR 2 \citep{Gaia2018}, the Two Micron All Sky Survey \citep{Cutri2003}, and the Wide-field Infrared Survey Explorer \citep{Cutri2014}. \texttt{EXOFASTv2} computes spectral energy distributions from MIST models and the Gaia parallax measurements and fits them to the archival photometry to infer the properties of the host star. We placed normal priors on $T_{\rm eff}$ and [Fe/H] based on the \texttt{SpecMatch} analysis of the high S/N template spectrum of \host\ (Section~\ref{sec:spec}). The width of the $T_{\rm eff}$ prior was inflated to 115~K (2\%) and a noise floor of 2\% was applied to bolometric flux used in the SED determination to account for systematic uncertainties inherent in the MIST models \citep{Tayar2020}. We also placed a uniform prior on extinction ($A_{\rm V}\le0.1376$) using the galactic dust maps of \citet{Schlafly2011} and a normal prior on parallax ($\varpi = 8.597\pm0.017$~mas) using the Gaia DR3 measurement corrected for the zero point offset \citep{Lindegren2021,Gaia2021}. These priors are summarized at the top of Table~\ref{tab:stellar}. 

\begin{deluxetable}{lcc}
\tabletypesize{\scriptsize}
\tablecaption{Median values and 68\% confidence intervals for the stellar parameters for \host. \label{tab:stellar}}
\tablehead{\colhead{~~~Parameter} & \colhead{Units} & \colhead{Values}}
\startdata
\multicolumn{2}{l}{Informative Priors:}& \smallskip\\
~~~~$T_{\rm eff}$  &Effective Temperature (K)  & $\mathcal{N}(5739,115)$\\
~~~~$[{\rm Fe/H}]$  &Metallicity (dex)  & $\mathcal{N}(0.25,0.06)$\\
~~~~$\varpi$  &Parallax (mas)  & $\mathcal{N}(8.597,0.017)$\\
~~~~$A_V$  &V-band extinction (mag)  & $\mathcal{U}(0,0.1376)$\\
\smallskip\\\multicolumn{2}{l}{Stellar Parameters:}&\smallskip\\
~~~~$M_*$  &Mass (\msun)  &$1.111^{+0.047}_{-0.046}$\\
~~~~$R_*$  &Radius (\rsun)  &$1.636^{+0.033}_{-0.029}$\\
~~~~$L_*$  &Luminosity (\lsun)  &$2.544^{+0.091}_{-0.093}$\\
~~~~$F_{Bol}$  &Bolometric Flux (cgs)  &$6.01\times10^{-9}$$^{+2.1\times10^{-10}}_{-2.2\times10^{-10}}$\\
~~~~$\rho_*$  &Density (g~cm$^{-3}$)  &$0.359^{+0.015}_{-0.016}$\\
~~~~$\log{g}$  &Surface gravity (cgs)  &$4.057^{+0.015}_{-0.016}$\\
~~~~$T_{\rm eff}$  &Effective Temperature (K)  &$5695^{+58}_{-60}$\\
~~~~$[{\rm Fe/H}]$  &Metallicity (dex)  &$0.253\pm0.057$\\
~~~~$[{\rm Fe/H}]_{0}$  &Initial Metallicity$^{a}$ (dex)   &$0.269^{+0.055}_{-0.054}$\\
~~~~${\rm Age}$  &Age (Gyr)  &$8.1^{+1.5}_{-1.3}$\\
~~~~$EEP$  &Equal Evolutionary Phase$^{b}$   &$452.1^{+3.9}_{-5.0}$\\
~~~~$A_V$  &V-band extinction (mag)  &$0.077^{+0.041}_{-0.048}$\\
~~~~$\sigma_{SED}$  &SED photometry error scaling   &$0.69^{+0.24}_{-0.16}$\\
~~~~$\varpi$  &Parallax (mas)  &$8.597\pm0.017$\\
~~~~$d$  &Distance (pc)  &$116.32\pm0.23$\\
\enddata
\tablenotetext{}{See Table~3 in \citet{Eastman2019} for a detailed description of all parameters and all default (non-informative) priors beyond those specified here. $\mathcal{N}(a,b)$ denotes a normal distribution with mean $a$ and variance $b^2$. $\mathcal{U}(a,b)$ denotes a uniform distribution over the interval [$a$,$b$].}
\tablenotetext{a}{Initial metallicity is that of the star when it formed.}
\tablenotetext{b}{Corresponds to static points in a star's evolutionary history. See Section~2 of \citet{Dotter2016}.}
\end{deluxetable}

We gauged convergence of the \texttt{EXOFASTv2} fit by the number of independent draws \citep{Ford2006b}, which exceeded 1,000, and the Gelman--Rubin statistic \citep{Gelman1992}, which was smaller than 1.01, for every fitted parameter.  From the fit parameters, numerous properties of \planet\ were derived. Table~\ref{tab:planet} lists all relevant planetary parameters for \planet. The derived planetary radius and transit depth assume no oblateness \citep[e.g.,][]{Seager2002} and no system of rings \citep[e.g.,][]{Barnes2004,Akinsanmi2018}, although \planet\ could plausibly have both.

\begin{deluxetable}{lcc}
\tabletypesize{\scriptsize}
\tablecaption{Median values and 68\% confidence interval of the planet parameters for \planet. \label{tab:planet}}
\tablehead{\colhead{~~~Parameter} & \colhead{Units} & \colhead{Values}}
\startdata
\multicolumn{2}{l}{Planetary Parameters:}&\smallskip\\
~~~~$P$  &Period$^{a}$ (d)  &$260.79^{+0.59}_{-0.58}$\\
~~~~$R_P$  &Radius (\rj)  &$1.010^{+0.022}_{-0.019}$\\
~~~~$M_P$  &Mass (\mj)  &$2.755^{+0.087}_{-0.081}$\\
~~~~$T_C$  &Time of conjunction (\bjdtdb)  &$2458830.7652\pm0.0010$\\
~~~~$a$  &Semi-major axis (au)  &$0.828\pm0.012$\\
~~~~$i$  &Inclination (degrees)  &$89.955^{+0.032}_{-0.044}$\\
~~~~$e$  &Eccentricity   &$0.3683\pm0.0073$\\
~~~~$\omega_*$  &Argument of Periastron (degrees)  &$-43.8\pm1.3$\\
~~~~$T_{eq}$  &Equilibrium temperature$^{b}$ (K)  &$348.0^{+3.3}_{-3.6}$\\
~~~~$\tau_{\rm circ}$  &Tidal circularization timescale (Gyr)  &$54600000^{+3800000}_{-4500000}$\\
~~~~$K$  &RV semi-amplitude (m~s$^{-1}$)  &$87.75^{+0.98}_{-0.99}$\\
~~~~$\dot \gamma$  &RV slope$^{c}$ (m~s$^{-1}$~day$^{-1}$)  &$-0.1205\pm0.0043$\\
~~~~$\ddot \gamma$  &RV quadratic term$^{c}$ (m~s$^{-1}$~day$^{-2}$)  &$0.000214^{+0.000039}_{-0.000038}$\\
~~~~$\delta_{\rm TESS}$  &Transit depth in TESS band  &$0.004766^{+0.000078}_{-0.000076}$\\
~~~~$\tau$  &Ingress/egress transit duration (d)  &$0.06044^{+0.0019}_{-0.00063}$\\
~~~~$T_{14}$  &Total transit duration (d)  &$1.0040^{+0.0032}_{-0.0031}$\\
~~~~$b$  &Transit Impact parameter   &$0.100^{+0.095}_{-0.070}$\\
~~~~$\rho_P$  &Density (g~cm$^{-3}$)  &$3.32^{+0.14}_{-0.16}$\\
~~~~$logg_P$  &Surface gravity (cgs)   &$3.827^{+0.012}_{-0.015}$\\
~~~~$\fave$  &Incident Flux (\fluxcgs)  &$0.00442\pm0.00018$\\
~~~~$T_P$  &Time of Periastron (\bjdtdb)  &$2458760.7\pm1.3$\\
~~~~$T_S$  &Time of eclipse (\bjdtdb)  &$2458745.41^{+1.00}_{-1.0}$\\
~~~~$b_S$  &Eclipse impact parameter   &$0.060^{+0.056}_{-0.042}$\\
~~~~$\tau_S$  &Ingress/egress eclipse duration (days)  &$0.03593^{+0.00098}_{-0.00088}$\\
~~~~$T_{S,14}$  &Total eclipse duration (days)  &$0.599\pm0.013$\\
~~~~$\delta_{S,2.5\mu m}$  &Blackbody eclipse depth at 2.5$\mu$m (ppm)  &$0.00237^{+0.00034}_{-0.00032}$\\
~~~~$\delta_{S,5.0\mu m}$  &Blackbody eclipse depth at 5.0$\mu$m (ppm)  &$1.54\pm0.10$\\
~~~~$\delta_{S,7.5\mu m}$  &Blackbody eclipse depth at 7.5$\mu$m (ppm)  &$11.29^{+0.49}_{-0.50}$\\
\smallskip\\\multicolumn{2}{l}{TESS Parameters:}& \smallskip\\
~~~~$u_{1}$  &linear limb-darkening coefficient   &$0.316\pm0.029$\\
~~~~$u_{2}$  &quadratic limb-darkening coefficient   &$0.248\pm0.044$\\
\smallskip\\\multicolumn{2}{l}{APF-Levy Parameters:}& \smallskip\\
~~~~$\gamma_{\rm rel}$  &Relative RV Offset$^{c}$ (m~s$^{-1}$)  &$-30.0^{+1.2}_{-1.3}$\\
~~~~$\sigma_J$  &RV Jitter (m~s$^{-1}$)  &$4.16^{+0.67}_{-0.62}$\\
\smallskip\\\multicolumn{2}{l}{Keck-HIRES Parameters:}& \smallskip\\
~~~~$\gamma_{\rm rel}$  &Relative RV Offset$^{c}$ (m~s$^{-1}$)  &$-12.7^{+1.4}_{-1.5}$\\
~~~~$\sigma_J$  &RV Jitter (m~s$^{-1}$)  &$4.86^{+1.3}_{-0.93}$\\
\enddata
\label{tab:HD238894.}
\tablenotetext{}{See Table 3 in \citet{Eastman2019} for a detailed description of all parameters and all default (non-informative) priors.}
\tablenotetext{a}{This orbital period is derived from the full posterior from the \texttt{EXOFASTv2} fit. See Section~\ref{sec:teers} for a description of the likely orbital period values ({$260.18^{+0.19}_{-0.30}$}~days or {$261.76^{+0.29}_{-0.16}$}~days) after the ground-based photometric transit recovery campaign for \host.}
\tablenotetext{b}{Assumes a Jupiter-like Bond albedo (0.34) and perfect heat redistribution.}
\tablenotetext{c}{Reference epoch = 2459157.439110}
\end{deluxetable}

The TESS data and the best fit transit model are shown in Figure~\ref{fig:tess}. All RV data and the best fit RV model are shown in Figure~\ref{fig:rv}. We included a quadratic function of time in the fit to account for the force from an additional planet or star on \host. Both coefficients in the quadratic function are greater than {$5\sigma$} discrepant from zero, suggesting that there is indeed a long-term variation in the RVs. Even with more than a {500}~day baseline of observations, we do not sample enough of the long-term signal to determine its cause. The lack of correlation between the RVs and the $S_{\rm HK}$ activity indicators disfavors stellar activity as the true explanation (Section~\ref{sec:spec}). Instead, we suggest that another massive object is orbiting \host\ (Section~\ref{sec:drift}).

Owing to the well sampled time series of precise RVs, the posterior for orbital period for the single-transit \planet\ has a standard deviation below 1~day ({0.23\%}) that is well characterized by a normal distribution. In the following section, we will further constrain the orbital period of \planet\ based on ground based photometry acquired near the timing of an additional transit.


\section{Ephemeris Refinement from Ground-based Observations}\label{sec:teers}

With each new RV we acquired of \host, we calculated the timing of the next transit of \planet. The second transit (i.e., the first transit to occur since the TESS single-transit) occurred at some point in late August or early September of 2020. At that time, the $2\sigma$ transit window---which was almost entirely determined by the uncertainty on orbital period---was approximately 10~days wide\footnote{Note that the uncertainty for orbital period reported in Table~\ref{tab:planet} is informed by the full RV data set and is therefore much smaller than the corresponding estimate in August 2020.}. 

We attempted the formidable task of observing this 24-hour-long, relatively shallow (0.5\%) transit by planning a global ground-based photometry campaign. In total, {15} telescopes acquired {55} data sets containing over {20,000} individual exposures of \host\ spanning 11~days. These included professional observatories, amateur observatories, and two portable digital {\it eVscope} telescopes. Each data set was processed with standard differential aperture photometry using background stars as references. Basic information about each telescope is provided in Appendix~\ref{app:ground}, a summary of each observation is listed in Table~\ref{tab:ground_obs}, and all data are plotted in Figure~\ref{fig:all_ground}.

None of the observations provided a conclusive detection of a \planet\ transit. As has been the case for other attempts to detect transits of long-period planets \citep[e.g.,][]{Winn2009,Dalba2016,Dalba2019c}, any single observation could only observe out-of-transit baseline and ingress or egress at best. This photometric signature can be easily mimicked by flux variations between the target and reference stars as the airmass changed. On the other hand, owing to the duration of the transit and the difficulty of absolute flux calibration at the precision of the transit depth, distinguishing a fully in-transit observation from a fully out-of-transit one is challenging. Also, the $\sim$0.5\% transit depth was on the order of the noise floor for many of the sites. All of these factors contributed to the nondetection. 

However, despite the lack of an obvious transit detection, we developed a straightforward method to refine the orbital period of \planet\ by simultaneously searching all data sets for times where ingress or egress \emph{did not} occur to high statistical significance. Ruling out these times rules out chunks of the orbital period posterior. Conversely, at times when we cannot determine if ingress or egress occurred---or if ingress or egress even appears to be favored by the data---we do not rule out those portions of the posterior.

We began with the relative light curves for each data set, which were the aperture flux values of \host\ divided by the aperture flux of one or more reference stars. We sigma-clipped the relative light curves for 4$\sigma$ outliers and normalized each to its median. So far, we had not attempted to remove flux variations due to airmass. Our first task was to remove data sets with high scatter to avoid introducing spurious results in the ephemeris refinement. Although even imprecise data contain useful information, we found that many of the high-scatter data sets contained time correlated noise or systematic noise features. Instead of attempting to correct for these noise properties, we chose to exclude the data set entirely. For the purpose of this procedure, we fit and subtracted an airmass model
\begin{equation}\label{eq:am}
    F^{\prime}(t) = c_1 e^{-c_2 X(t)}
\end{equation}

\noindent where $F^{\prime}$ is the flux variation owing to a varying airmass $X$, both of which are functions of time $t$. The model had two fitted coefficients $c_1$ and $c_2$, which can have any finite value. After this airmass correction, data sets for which the standard deviation exceeded the transit depth (0.5\%) were removed (gray points in Figure~\ref{fig:all_ground}). We list this standard deviation for each data set in Table~\ref{tab:ground_obs}.

Next, we established a fine linear grid of mid-transit ($T_0$) times that spanned fourth contact at the time of the very first flux measurement (BJD$_{\rm TDB} = 2459085.674949$) to first contact at the time of our very last flux measurement (BJD$_{\rm TDB} = 2459096.875975$). These $T_0$ values were used to generate transit models that were compared with the data. Each value of $T_0$ maps to a unique value of orbital period.

Then, we iterated over each $T_0$ value and each data set conducting the following procedure. A transit model was generated at that $T_0$ using the \texttt{batman} package \citep{Kreidberg2015b}. All other transit parameters were fixed at the values derived from the \texttt{EXOFASTv2} model (Table~\ref{tab:planet}) except for the quadratic limb darkening coefficients, which were drawn from the look-up tables of \citet{Claret2011} according to filter. We subtracted this transit model from the relative light curve that was sigma-clipped and normalized to its median. Note that the light curve in this case had not been corrected for airmass variations, which were therefore still present after the subtraction. The model-subtracted light curve was then fit to Equation~\ref{eq:am} with a basic least-squares algorithm that minimized the $\chi^2$ statistic. 

Consider the logic behind this procedure. If the airmass model is a good fit to the light curve after the transit model is subtracted (i.e., low $\chi^2$ value), then we failed to rule out the transit at that time. We also cannot confidently claim that we have detected the transit, since the good fit to the airmass model could be coincidental. In this case, the probability of this transit time is still described by the posterior from the RV and transit joint fit. Conversely, if the airmass model is a poor fit to the light curve after the transit model is subtracted (i.e., high $\chi^2$ value), then we can be confident that the transit did not occur at that time.

Each $T_0$ for each data set yielded its own minimum $\chi^2$ value from which we calculated the corresponding log likelihood value (i.e., the maximum of the likelihood function) and Bayesian information criterion \citep[BIC;][]{Schwarz1978}. We then summed the BIC values across data sets to produce a total BIC as a function of $T_0$ and hence orbital period. 

We also repeated our entire procedure and calculated the same metrics assuming that none of the observations sampled the transit. In this ``no transit'' scenario, the transit occurred either before or after the observations or fell entirely within a data gap.  

\begin{figure}
    \centering
    \includegraphics[width=\columnwidth]{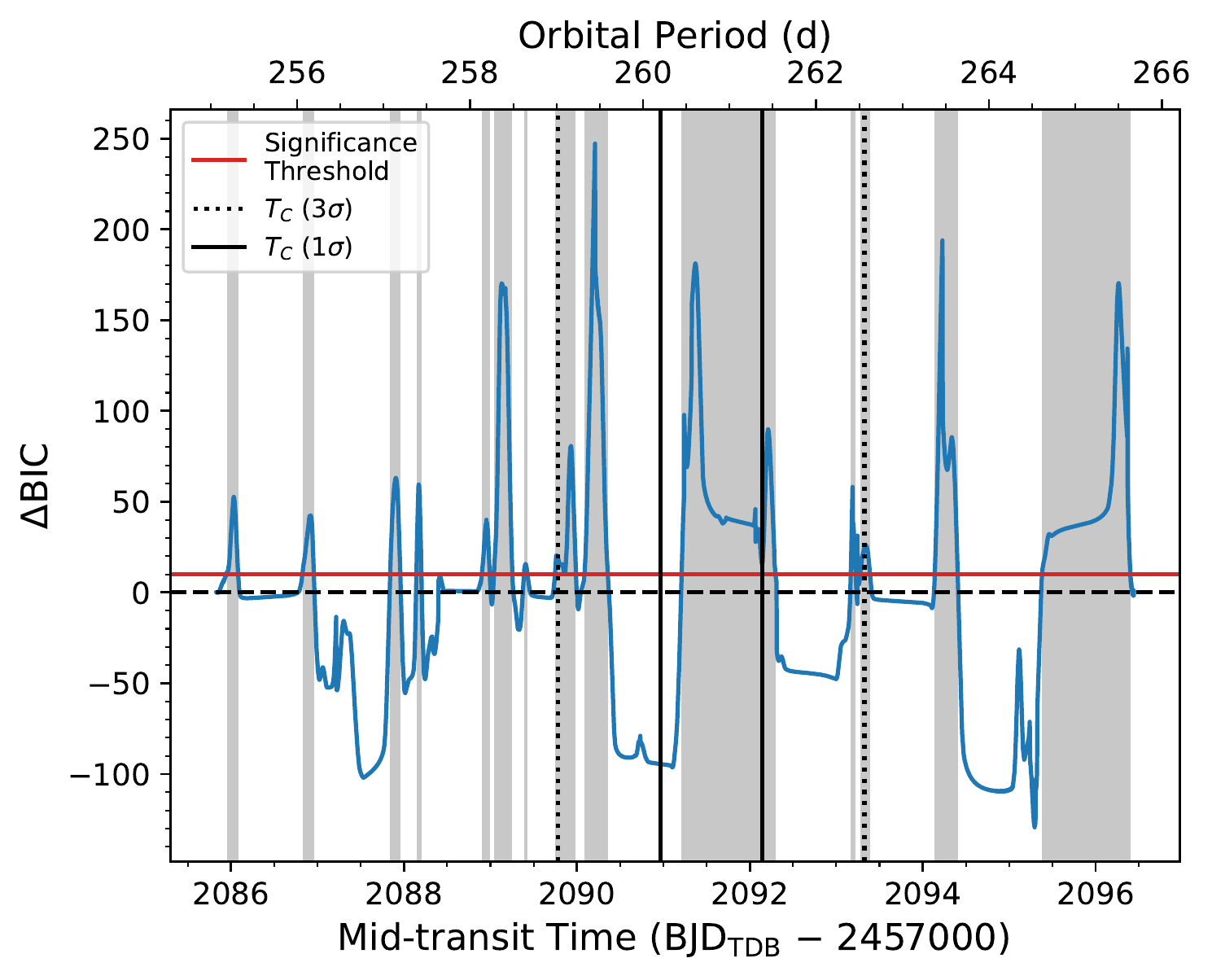}
    \caption{Summary of the transit ephemeris refinement analysis. Positive values of $\Delta$BIC indicate mid-transit times that are disfavored relative to a nondetection. We consider mid-transit times with $\Delta$BIC$\ge10$ (gray regions) to be ruled out by the ground-based photometry. Negative values of $\Delta$BIC indicate times where the transit model provided a good fit. However, this could be coincidental and we do not interpret large negative $\Delta$BIC as evidence of the transit. The $1\sigma$ and $3\sigma$ ranges for mid-transit time ($T_C$) from the final ephemeris (Table~\ref{tab:planet}) are shown as vertical lines. Each mid-transit time corresponds to a specific orbital period of \planet, as indicated at the top of the figure.}
    \label{fig:teers}
\end{figure}

\begin{figure}
    \centering
    \includegraphics[width=\columnwidth]{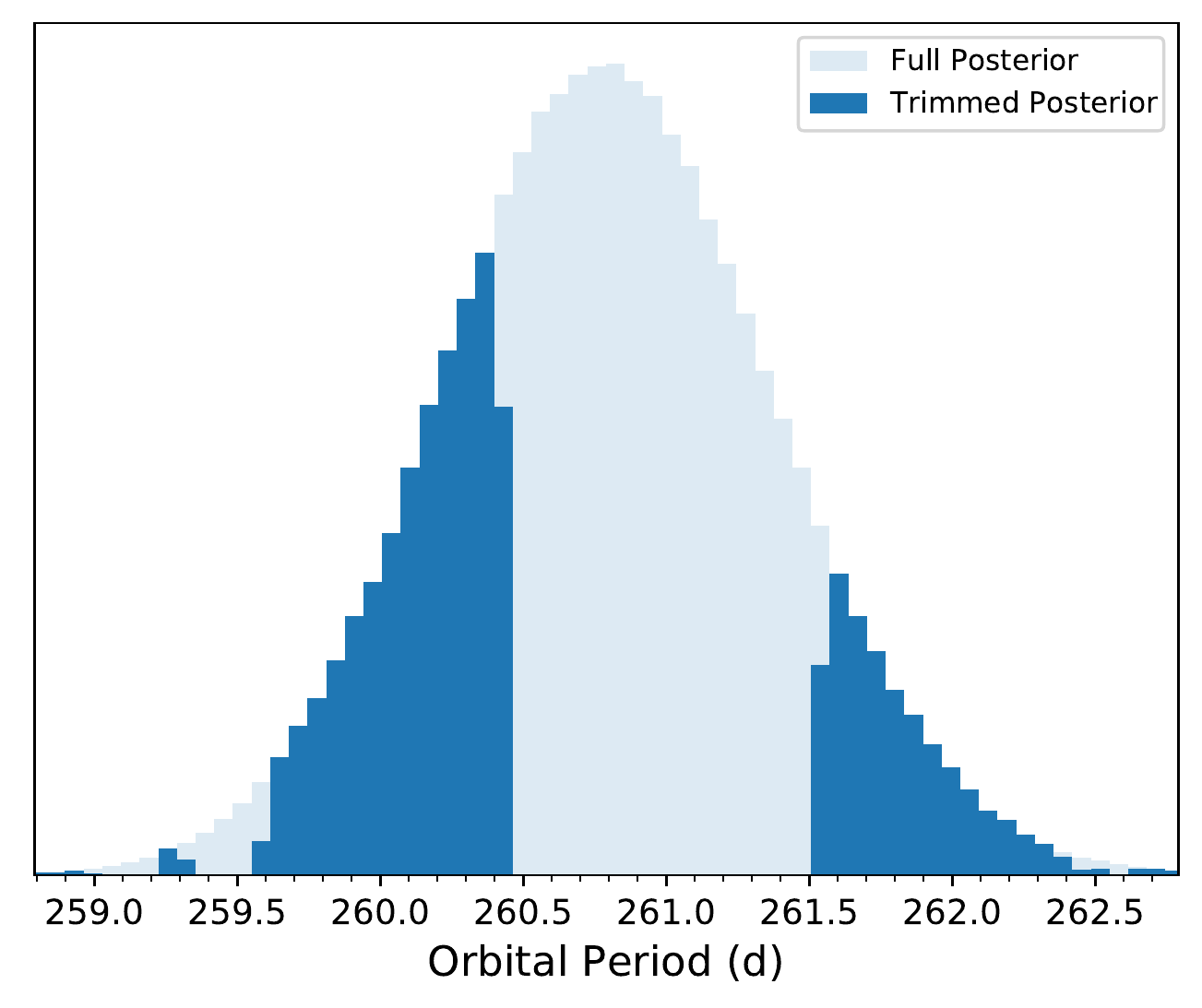}
    \caption{Orbital period posterior for \planet. The light blue shows the full posterior while the dark blue shows the region that is still allowed after the transit ephemeris refinement using the ground-based photometry, which rules out the most likely orbital periods (around the mean and median) inferred from the comprehensive \texttt{EXOFASTv2} analysis.}
    \label{fig:trimmed_posterior}
\end{figure}

In Figure~\ref{fig:teers}, we show the difference in BIC values ($\Delta$BIC) between the transit and ``no transit'' scenarios as a function of mid-transit time. The earliest and latest values of $\Delta$BIC approach zero as expected. Values of $T_0$ with substantially negative $\Delta$BIC primarily correspond to times when ingress and/or egress occurred during data gaps and the relatively flat ground-based photometry mimics the basin of a transit. On the other hand, values of $T_0$ with substantially positive $\Delta$BIC correspond to ingress and/or egress lining up with ground-based photometry that clearly does not contain such features. We adopted a $\Delta$BIC threshold of 10---corresponding to ``very strong`` evidence against the transit model \citep{Kass1995}---to determine which values of $T_0$ we could rule out (Figure~\ref{fig:teers}, gray regions) and mapped this refinement back to orbital period. The resulting trimmed posterior for orbital period is shown in Figure~\ref{fig:trimmed_posterior}.

The ground-based transit detection campaign served to broadly divide the normal posterior for orbital period into two smaller, non-Gaussian groups while ruling out the median and most likely values of the original distribution. Described by their median and 68\% credible intervals, the two possible orbital periods are {$260.18^{+0.19}_{-0.30}$}~days and {$261.76^{+0.29}_{-0.16}$}~days.

To assess the efficiency of the ground-based photometry campaign, we consider the duty cycle between exposure time and orbital period posterior time ruled out. Multiplying the number of exposures by their respective exposure times (Table~\ref{tab:ground_obs}) for all observations (regardless of whether they were excluded from this analysis) yields {4.4}~days. The aforementioned $\Delta$BIC threshold of 10 rules out {3.8}~days worth of orbital period posterior space. Therefore, the duty cycle of the campaign was {86\%} (i.e., for every 1~hour of exposure time, we ruled out {52~minutes} of orbital period). Ideally, a campaign like this would detect the transit and the duty cycle would be less important. However, this metric can be recalculated for any future single-transit detection campaigns that yield nondetections or proposals to conduct such campaigns to compare strategies and assess effectiveness.

\subsection{Prospects for Future Transit Detection}\label{sec:future_tran}

Despite the ephemeris refinement for \planet\ conducted here, future characterization of this planet will be challenging until another transit is detected. The third transit of \planet\ (assuming the TESS transit as the first) occurred sometime in mid-May 2021 and was not observed. The fourth transit is predicted to occur in late January or early February 2022. Specifically, our predictions following the trimmed orbital period posterior are {19:19:30 UTC 31 January 2022} and {13:05:05 UTC 5 February 2022}. The uncertainties on these predictions are asymmetric but roughly on the order of {a day} or less ($1\sigma$). 

Fortunately, TESS is expected to re-observe \host\ in Sector~48 of the extended mission beginning around 28 January 2022\footnote{According to the Web TESS Viewing Tool (\url{https://heasarc.gsfc.nasa.gov/cgi-bin/tess/webtess/wtv.py}) accessed 2021 September 4.}. We therefore predict that TESS will observe another transit of \planet\ at that time. Barring data gaps, if another transit is not seen by TESS in Sector 48, then only a small (relatively unlikely) tail of the orbital period posterior distribution would be consistent with the original single-transit and the RVs.  

A second transit detection would drastically reduce the uncertainty on the orbital period and preserve the transit ephemeris for years into the future. However, some giant planets on 100--1,000~day orbits are known to exhibit day-long timing variations from transit-to-transit \citep[e.g.,][]{Wang2015b}. A third transit would need to be observed to explore the existence of transit timing variations \citep[e.g.,][]{Dalba2019c}.


\section{Bulk Heavy-element Analysis}\label{sec:metal}

\planet\ is one of a small but growing collection of valuable giant transiting exoplanets on $\sim$au-scale orbits with precisely measured masses and radii \citep[e.g.,][]{Dubber2019,Chachan2021,Dalba2021a,Dalba2021c}. With these two properties, we can infer bulk heavy-element mass and metallicity relative to stellar to better understand their structure and formation history. 

Following \citet{Thorngren2019a}, we generated one-dimensional spherically symmetric giant planet structure models with a rock/ice core, a convective envelope consisting of rock, ice, and H/He, and a radiative atmosphere interpolated from the \citet{Fortney2007} grid. Along with the mass, radius, and age for \planet\ (drawn from the posteriors of the \texttt{EXOFASTv2} fit), the models recovered the total mass of heavy elements and thereby bulk metallicity needed to explain the planet's size. We find that the bulk metallicity for \planet\ is {$Z_p = 0.12\pm0.03$}, which corresponds to {105~$M_{\earth}$} of heavy elements. We can approximate the stellar metallicity using the iron abundance following $Z_{\star} = 0.142 \times 10^{\rm [Fe/H]}$, which gives {$Z_{\star} = 0.0254\pm0.0033$.} This sets the metal enrichment ($Z_p/Z_{\star}$) at {$4.7\pm1.3$}. 

The metallicity enrichment of \planet\ is consistent with the core accretion theory of giant planet formation \citep{Pollack1996} with late-stage accretion of icy planetesimals, which can explain enrichment by a factor of a few to a dozen \citep[e.g.,][]{Gautier2001,Mousis2009}. An alternate theory to late stage accretion---the mergers of planetary cores during the gas accretion phase \citep[e.g.,][]{Ginzburg2020}---can explain enrichment factors between 1.5--10 for a {2.7~$M_{\rm J}$} planet, making it a viable formation pathway as well.

\begin{figure}
    \centering
    \includegraphics[width=\columnwidth]{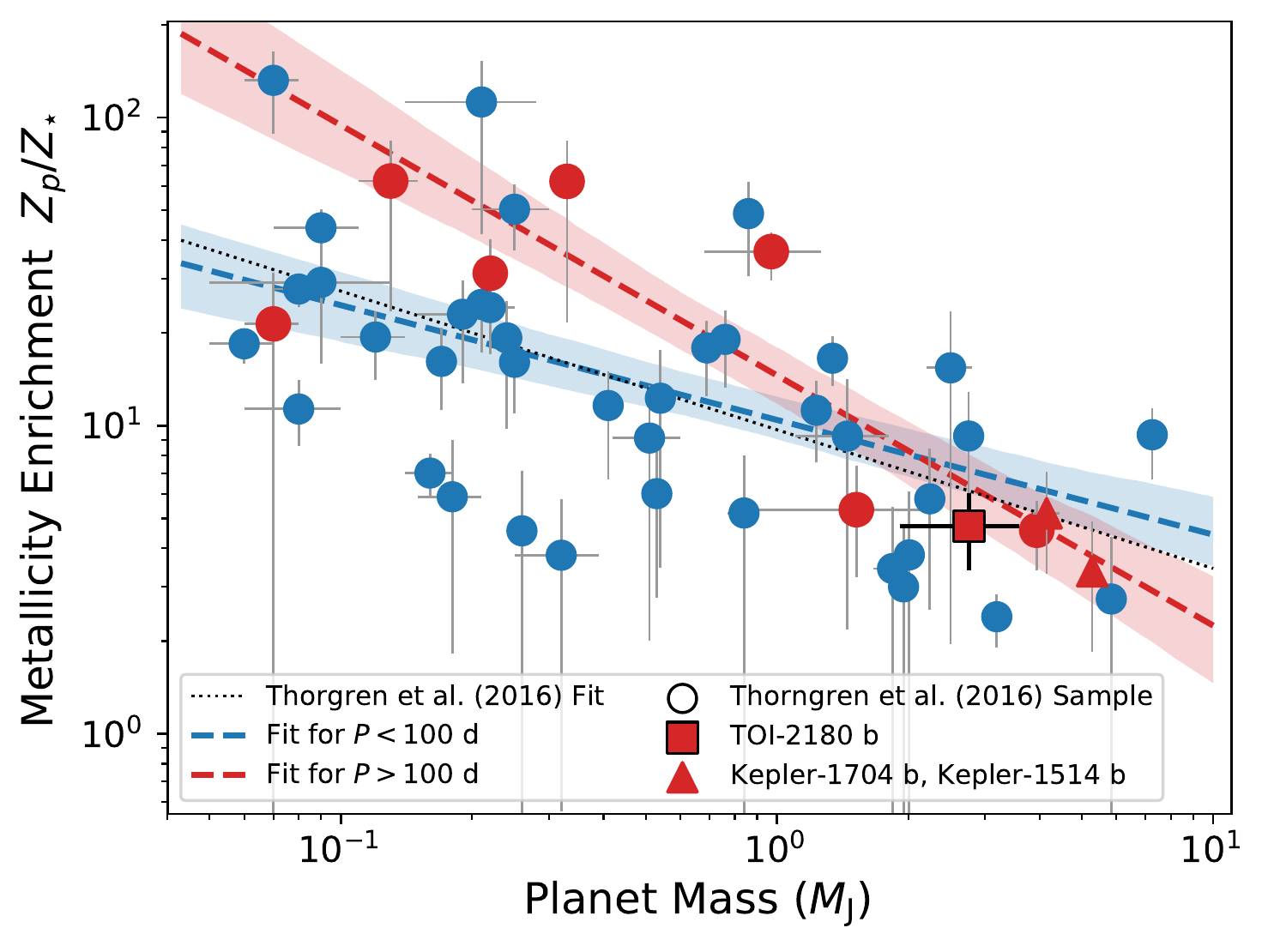}
    \caption{Giant planet mass--metallicity correlation shown with the \citet{Thorngren2016} sample (circles), \planet\ (square), and two recently published \kepler\ long-period giant planets (triangles): Kepler-1514~b ($P\approx218$~days) and Kepler-1704~b ($P\approx989$~days). Blue and red indicate orbital periods below and above 100~days, respectively. The dashed lines and shaded $1\sigma$ regions indicate separate regressions to these two sets of planets. The slopes of these fits are discrepant to $2.6\sigma$, hinting that the mass--metallicity correlation for giant planets may be dependent upon orbital properties.}
    \label{fig:zpzs}
\end{figure}

In Figure~\ref{fig:zpzs}, we plot the mass and metal enrichment of \planet\ relative to the \citet{Thorngren2016} sample of giant exoplanets. We also include two other recently published high mass, giant exoplanets on long-period orbits: Kepler-1514~b \citep[$P\approx218$~days;][]{Dalba2021a} and Kepler-1704~b \citep[$P\approx989$~days;][]{Dalba2021c}. The metal enrichment of \planet\ relative to its host star is consistent with other giant planets with similar mass and falls near to the best fit line for all of the \citet{Thorngren2016} sample. 

The \citet{Thorngren2016} exoplanets are plotted in Figure~\ref{fig:zpzs} as circles. The points are given different colors based on orbital period. The two \kepler\ planets are shown as triangles and \planet\ is shown as a square. Although each of these planets individually is fully consistent with the \citet{Thorngren2016} mass--metallicity correlation, there is possibly a subtle difference in the slope of the relation for the longer-period planets. For lower mass planets, enrichment appears to be higher than average for the longer-period objects and vice versa for the higher mass planets including \planet. 

We explored this possibility quantitatively by separating the planets in Figure~\ref{fig:zpzs} into short-period ($P<100$~days) and long-period ($P\ge100$~days) groups. The median orbital periods in the two groups were 10~days and 223~days. \planet\ and the \kepler\ planets are also given the corresponding $P>100$~days color. Although 100~days is a somewhat arbitrary separation value, it possibly separates planets that experienced different formation histories. We conducted orthogonal distance regression fits, which include uncertainties on both the explanatory and response variables, to both sets of planet masses and metal enrichments in logspace. The resulting power-law fits are drawn as dashed lines in Figure~\ref{fig:zpzs}. The $1\sigma$ uncertainty regions are also shown. The fits for the short and long-period planets are ($10.4\pm1.6$)$M^{(-0.372\pm0.084)}$ and (14.6$\pm$2.9)$M^{(-0.81\pm0.14)}$, respectively. The slopes in these fits are inconsistent at $2.6\sigma$. 

The mass--metallicity correlation derived for planets with orbital periods below 100~days is fully consistent with that measured by \citet{Thorngren2016}. On the other hand, the small set of long-period planets that includes \planet\ produces a notably steeper correlation. We refrain from placing too much emphasis on this finding owing to the small number of data points and moderate statistical significance. The addition of any number of additional long-period giant planets would be elucidating. If this trend is real, though, it suggests that the current orbital properties of giant planets trace different heavy-element accretion mechanisms such as pebble or planetesimal \citep{Hasegawa2018}. A statistically robust analysis of the mass--metallicity correlation is warranted, but we leave such an analysis to future work.

For a solar system comparison, the {\it Galileo Entry Probe} measured volatile gases in Jupiter's atmosphere and identified enrichment of 2--6 for several heavy elements and noble gasses \citep{Wong2004}. More recently, the {\it Juno} spacecraft measured the equatorial water abundance on Jupiter to be 1--5 times the protosolar value \citep{Li2020}. The comparison between our enrichment measurement of \planet\ and these measurements at Jupiter comes with caveats. For instance, the Jupiter enrichment is derived from its equatorial oxygen abundance while the exoplanet enrichment is a model dependent bulk value. The direct comparison of these two qualities may be problematic. However, the main point is that these values are all of a similar order of magnitude.


\section{Analysis of RV Drift}\label{sec:drift}

We characterized the trend and curvature in TOI-2180's RV time series using the technique described in \citet{Lubin2021}. Because TOI-2180 is not in the \textit{Hipparcos} catalog \citep{ESA1997}, we could not calculate its astrometric acceleration using the \textit{Hipparcos--Gaia} Catalog of Accelerations \citep{Brandt2018}. Instead we analyzed only the RV data, which leaves a degeneracy between companion mass and semi-major axis.

\begin{figure}
    \centering
    \includegraphics[width=\columnwidth]{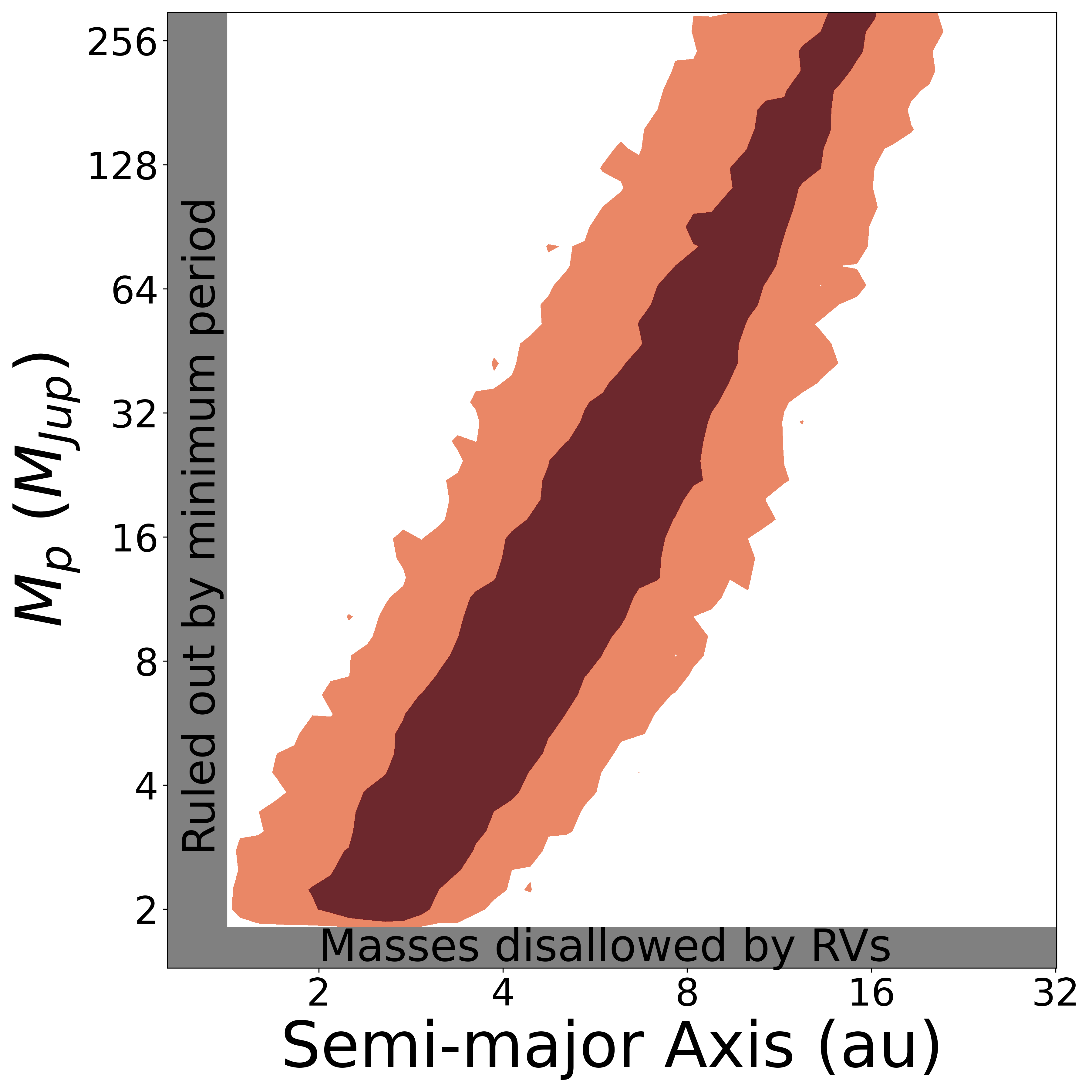}
    \caption{The set of $(a, M_P)$ models consistent with the measured RV trend and curvature at the 1$\sigma$ (dark) and 2$\sigma$ (light) levels. We rule out semi-major axes corresponding to orbital periods shorter than the observing baseline, as well as companion masses too low to produce the minimum RV amplitude.}
    \label{fig:RV_drift}
\end{figure}

We sampled $10^8$ synthetic companion models, each comprising a mass $M_P$, semi-major axis $a$, eccentricity $e$, inclination $i$, argument of periastron $\omega$, and mean anomaly $M$. For each model companion, we calculated the trend ($\dot{\gamma}$) and curvature ($\ddot{\gamma}$) that such a companion would produce. We then calculated the model likelihood given the measured trend and curvature in Table \ref{tab:planet}, and marginalized over $\{e, i, \omega, M\}$ by binning the likelihoods in $a$-$M_P$ space. Figure \ref{fig:RV_drift} shows the joint posterior distribution for TOI-2180's companion.

If located within a few au, the object is likely to be planetary and more massive than $\sim$2~$M_{\rm J}$. Past roughly 8~au, however, the object transitions to the substellar regime. We truncate the upper mass boundary of Figure~\ref{fig:RV_drift} at a few tenths of a solar mass at which point we would have expected to detect such a companion in the speckle imaging (Section~\ref{sec:imaging}). For a 13~$M_{\rm J}$ object, if we extend the RV monitoring by an additional $\sim$3.5~years, we will have sampled between 25\% and 100\% of the full orbit by period. At that point, we should be more capable of distinguishing between planetary and non-planetary scenarios.


\section{Discussion}\label{sec:disc}

With an equilibrium temperature of {348 K} (assuming a Jupiter-like Bond albedo, see Table~\ref{tab:planet}), \planet\ qualifies as a temperate Jupiter that occupies an interesting region of parameter space. It exists within 1--3~au where the giant planet occurrence rate increases \citep[e.g.,][]{Wittenmyer2020,Fernandes2019,Fulton2021}, but it is not so close to its star that its orbit has been tidally circularized (e.g., following a high eccentricity migration pathway). This means that its orbital properties may contain information about previous migration. \planet\ is warmer than Jupiter and Saturn, but it receives a weak enough irradiation to not be inflated. Planets like this are useful laboratories for models of interior and atmospheric structure and planet formation. 

\begin{figure}
    \centering
    \includegraphics[width=\columnwidth]{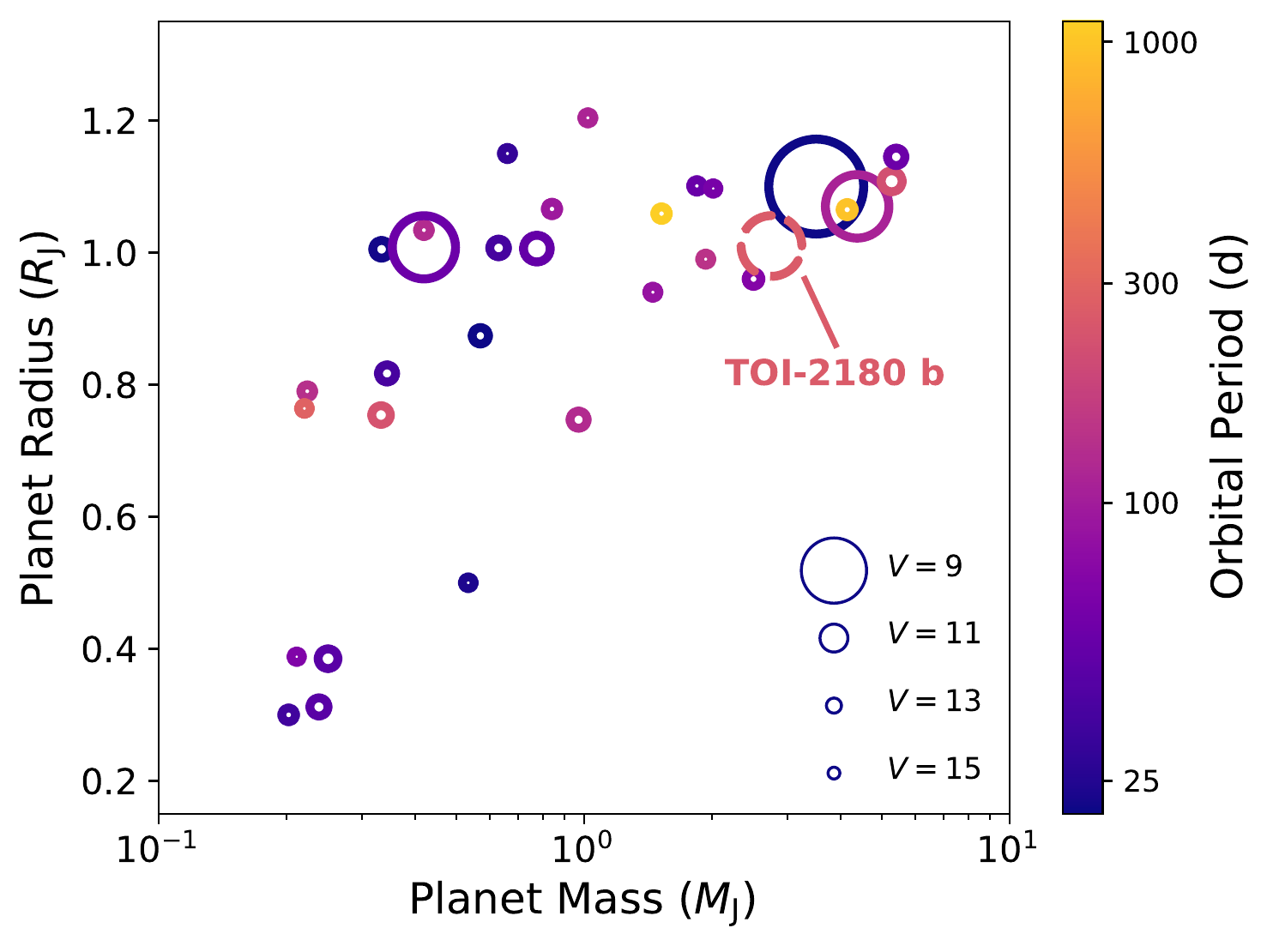}
    \caption{The subset of long-period ($P>20$~days), giant ($M_P>0.2$~$M_{\rm J}$) exoplanets with measured mass and radii sized to show their hosts' $V$ band magnitudes. \planet\ is labeled and drawn with a dashed line. Of the four planets with the brightest host stars, \planet\ has the longest period by far demonstrating how it is a valuable extension of the parameter space of temperate Jupiters.}
    \label{fig:MR}
\end{figure}

\planet\ stands out among other transiting temperate Jupiters for two main reasons: its long orbital period and its host star's favorable brightness. In Figure~\ref{fig:MR}, we put \planet\ in context with other giant planets ($M_P>0.2$~$M_{\rm J}$) with orbital periods greater than 20~days (to exclude hot Jupiters) with measured mass and radius that did not have a controversial flag\footnote{According to the NASA Exoplanet Archive, accessed 2021 September 5}. The \kepler\ mission has so-far proven most successful at discovering planets in this region of parameter space \citep[e.g.,][]{Wang2015b,Uehara2016,ForemanMackey2016b,Kawahara2019}. As a result, many of the planets in Figure~\ref{fig:MR} are sufficiently faint that follow-up opportunities to further characterize their systems are very limited. TESS, however, is slowly beginning to populate the long-period giant planet parameter space with systems orbiting bright host stars that are more amenable to follow-up characterization \citep[e.g.,][]{Eisner2020}.

As a temperate giant planet, the atmosphere of \planet\ represents a stepping stone between well characterized hot Jupiters and the cold solar system gas giant planets. At {348~K}, we might expect to find non-equilibrium carbon chemistry that relies heavily on vertical transport \citep{Fortney2020}. Moreover, an atmospheric metallicity measurement of \planet\ would be an ideal comparison to the now known stellar and bulk planetary metallicity. Perhaps the atmospheric metallicity will be lower than the bulk value since some amount of heavy elements are likely sequestered into a core. However, evidence from solar system observations suggests that giant planets may instead have interior regions of inward-decreasing metallicity \citep[e.g.,][]{Wahl2017,Guillot2018,Debras2019}. These theories could possibly be explored with transmission spectroscopy, which is suspected to probe CH$_4$ and the byproducts of disequilibrium chemistry and photolysis in long-period giant planets \citep{Dalba2015,Fortney2020}. However, owing to its high surface gravity and an unfavorable planet-star radius ratio, \planet\ is likely not amenable to transmission spectroscopy. The atmospheric scale height of \planet\ is only {$\sim$23~km}, which corresponds to a transit depth of a few parts per million (PPM) and a transmission spectroscopy metric of {7} \citep{Kempton2018}. Similarly, the near-infrared depth of the secondary eclipse is likely to be on the order of or less than 10~ppm (Table~\ref{tab:planet}). Any future endeavor to identify favorable targets for temperate Jupiter atmospheric characterization should examine whether the brightness of \host\ compensates for the difficulty introduced by its large stellar radius and the high surface gravity of \planet.  

Other avenues of follow-up characterization for the \host\ system are more promising. Continued RV monitoring of \host\ would eventually capture a sufficient fraction of the long-term acceleration to infer the properties of the outer companion. This RV trend could also be interpreted jointly with imaging and astrometric data to further constrain the outer object's orbit and inclination \citep[e.g.,][]{Crepp2012,Wittrock2016,Kane2019b,Brandt2021,Dalba2021b}.  Be it a star or substellar object, it could have influenced the evolution and migration of \planet, including eccentricity excitation through Kozai-Lidov oscillations \citep[e.g.,][]{Wu2003,Fabrycky2007,Naoz2012}. However, we need not invoke secular interactions \citep[e.g.,][]{Wu2011} or planet-planet scattering \citep[e.g.,][]{Rasio1996} to explain the moderate eccentricity ({$e\approx0.37$}) of \planet. \citet{Debras2021} argued that disk cavity migration could explain warm Jupiter eccentricity up to $\sim$0.4. Additional modeling of the formation and dynamical evolution of the objects in the \host\ system, with comparison to the bulk metallicity of \planet, would be illuminating.

The degeneracy between disk migration and secular interactions could possibly be broken with a measurement of the stellar obliquity via the Rossiter-McLaughlin (RM) effect \citep{Rossiter1924,McLaughlin1924}. Migration via interactions with a distant massive object would cause both orbits to be misaligned with the stellar spin. Alternatively, if it migrated in the disk and the disk is assumed to have been aligned with the stellar equator, we would expect little obliquity. This assumption may be problematic, though, as disks can be tilted such that they yield misaligned planets under disk migration \citep[e.g.,][]{Spalding2016b}. Nonetheless, the \texttt{SpecMatch} analysis of the \host\ spectrum returned a low rotational velocity of $v\sin{i} = 2.2\pm1.0$~km~s$^{-1}$. The largest possible amplitude of the RM effect would therefore be $\sim$9~m~s$^{-1}$ \citep{Winn2010b}, which is well within the capabilities of next-generation precise RV facilities such as MAROON-X \citep{Seifahrt2018} or the Keck Planet Finder \citep{Gibson2016}. The RM effect would also be detectable from either Keck-HIRES or APF-Levy, which have average internal RV precisions of 1.2~m~s$^{-1}$ and 3.6~m~s$^{-1}$, respectively. Achieving the proper timing for such an experiment given the 24~hr transit is challenging, though, and may not be feasible for several years. The scientific benefit of an RM detection for this system, and for such a long-period planet in general, further motivates the need to refine the transit ephemeris.

Lastly, we briefly consider \planet\ as a host for exomoons. The TESS transit offered no evidence to suggest that an exomoon is present, but the stellar brightness, the long-period planetary orbit, and the (possibly) gentle migration history warrant raise the possibility of exomoon detection in this system. \host\ is sufficiently bright that the precision of a single-transit observation would likely reach the noise floors of NIRISS and NIRSpec \citep[several tens of PPM;][]{Greene2016,Batalha2017b}. A Ganymede-size moon would produce a $\sim$5~PPM transit, which would not be detectable. Alternatively, an Earth-size moon would yield a $\sim$30~PPM occultation, which is more reasonable. As more transits of \planet\ are observed by TESS or any other facility, we recommend that the community conduct photodynamical modeling to test for variations in transit timing and duration that might indicate the presence or lack of an exomoon \citep[e.g.,][]{kipping2012b,Heller2014b,Kipping2021}. 


\section{Summary}\label{sec:concl}

Single-transit events are the primary avenue to discovering exoplanets with orbital periods longer than approximately a month in TESS photometry. Here, we describe the follow-up effort surrounding a 24~hr long single-transit of \host\ (Figure~\ref{fig:tess}), a slightly evolved mid G type star, in Sector 19 data from TESS (Section \ref{sec:tess}). Citizen scientists identified the transit event shortly after the data became public, allowing a Doppler monitoring campaign with the APF telescope to begin immediately (Section~\ref{sec:spec}). After nearly two years of RV observations with the APF and Keck telescopes (Figure~\ref{fig:rv}), we determined that \planet---a {2.8}~$M_{\rm J}$ giant planet on a {260.79}~day, eccentric ({0.368}) orbit---was the cause of the single-transit event (Section~\ref{sec:tran_rv}). 

\planet\ is a member of a rare but growing sample of valuable \emph{transiting} giant exoplanets with orbital periods in the hundreds of days. We conduct a thorough comprehensive fit to the transit and RV data to infer the stellar and planetary properties of this system (Section~\ref{sec:EFv2}). Our precise and regularly sampled RVs refine the ephemeris of \planet, and we attempt to detect a second transit through 11~days of photometric observations with ground-based telescopes situated over three continents (Figure~\ref{fig:all_ground}). Although we do not detect a transit in these data, we develop a straightforward method to combine the orbital period posterior from the fit to the single-transit and RVs with the extensive collection of ground-based data sets (Section~\ref{sec:teers}). This analysis substantially refines the orbital period of \planet\ by eliminating a substantial fraction of the most likely posterior solutions (Figure~\ref{fig:trimmed_posterior}), leaving the prediction that TESS will likely detect the transit of \planet\ in January or February 2022 (Section~\ref{sec:future_tran}).     

With a measured mass and radius for \planet, we infer the bulk heavy-element content and metallicity relative to stellar from interior structure models (Section~\ref{sec:metal}). \planet\ likely has {over 100~$M_{\earth}$} of heavy elements in its envelope and interior and is enriched relative to its host star by a factor of {$4.7\pm1.3$}. We place \planet\ in context of the mass--metallicity correlation for giant planets in Figure~\ref{fig:zpzs}. Along with a few other recently characterized exoplanets on several hundred day orbital periods, \planet\ suggests at 2.6$\sigma$ confidence that the relation between metal enrichment (relative to stellar) and mass for giant planets is dependent on orbital properties. We leave further analysis of this possibility to future work.

Lastly, we place the discovery of \planet\ in context of other temperate giant planets with known mass and radius (Section~\ref{sec:disc}, Figure~\ref{fig:MR}). It is a poor candidate for transmission spectroscopy owing to its high surface gravity and the {1.6~$R_{\sun}$} radius of its host star. Still, this system is a promising target for continued RV monitoring and stellar obliquity measurement to test theories of how giant planets migrate within the occurrence rate increase near 1~au but not so close as to the become hot Jupiters. \planet\ also remains as an excellent candidate for exomoon investigations since the host star brightness ($V=9.2$; $J=8.0$) makes it amenable to incredibly precise space-based photometry in the future. \\ \\


\section*{Acknowledgements}
The authors recognize and acknowledge the cultural role and reverence that the summit of Maunakea has within the indigenous Hawaiian community. We are deeply grateful to have the opportunity to conduct observations from this mountain. 

The authors would like to thank the anonymous referee for helpful comments that improved this paper. We also thank Jack Lissauer for interesting conversations about giant planets that guided some of our analysis. We thank Ken and Gloria Levy, who supported the construction of the Levy Spectrometer on the Automated Planet Finder. We thank the University of California and Google for supporting Lick Observatory and the UCO staff for their dedicated work scheduling and operating the telescopes of Lick Observatory. Some of the data presented herein were obtained at the W. M. Keck Observatory, which is operated as a scientific partnership among the California Institute of Technology, the University of California, and NASA. The Observatory was made possible by the generous financial support of the W. M. Keck Foundation.

This paper includes data collected by the \tess\ mission. Funding for the \tess\ mission is provided by the NASA's Science Mission Directorate. We acknowledge the use of public TESS data from pipelines at the TESS Science Office and at the TESS Science Processing Operations Center. Resources supporting this work were provided by the NASA High-End Computing (HEC) Program through the NASA Advanced Supercomputing (NAS) Division at Ames Research Center for the production of the SPOC data products. This research has made use of the NASA Exoplanet Archive, which is operated by the California Institute of Technology, under contract with the National Aeronautics and Space Administration under the Exoplanet Exploration Program. This research has made use of the Exoplanet Follow-up Observation Program website, which is operated by the California Institute of Technology, under contract with the National Aeronautics and Space Administration under the Exoplanet Exploration Program. This paper includes data collected with the TESS mission, obtained from the MAST data archive at the Space Telescope Science Institute (STScI). STScI is operated by the Association of Universities for Research in Astronomy, Inc., under NASA contract NAS 5-26555. We acknowledge the use of public TESS data from pipelines at the TESS Science Office and at the TESS Science Processing Operations Center. This work makes use of observations from the LCOGT network. Part of the LCOGT telescope time was granted by NOIRLab through the Mid-Scale Innovations Program (MSIP). MSIP is funded by NSF.

P. D. is supported by a National Science Foundation (NSF) Astronomy and Astrophysics Postdoctoral Fellowship under award AST-1903811. E.A.P. acknowledges the support of the Alfred P. Sloan Foundation. L.M.W. is supported by the Beatrice Watson Parrent Fellowship and NASA ADAP Grant 80NSSC19K0597. A.C. is supported by the NSF Graduate Research Fellowship, grant No. DGE 1842402. D.H. acknowledges support from the Alfred P. Sloan Foundation, the National Aeronautics and Space Administration (80NSSC21K0652), and the National Science Foundation (AST-1717000). I.J.M.C. acknowledges support from the NSF through grant AST-1824644. R.A.R. is supported by the NSF Graduate Research Fellowship, grant No. DGE 1745301. C. D. D. acknowledges the support of the Hellman Family Faculty Fund, the Alfred P. Sloan Foundation, the David \& Lucile Packard Foundation, and the National Aeronautics and Space Administration via the TESS Guest Investigator Program (80NSSC18K1583). J.M.A.M. is supported by the NSF Graduate Research Fellowship, grant No. DGE-1842400. J.M.A.M. also acknowledges the LSSTC Data Science Fellowship Program, which is funded by LSSTC, NSF Cybertraining Grant No. 1829740, the Brinson Foundation, and the Moore Foundation; his participation in the program has benefited this work. T.F. acknowledges support from the University of California President's Postdoctoral Fellowship Program. N.E. acknowledges support from the UK Science and Technology Facilities Council (STFC) under grant code ST/R505006/1. D. D. acknowledges support from the TESS Guest Investigator Program grant 80NSSC19K1727 and NASA Exoplanet Research Program grant 18-2XRP18\_2-0136. M. R. is supported by the National Science Foundation Graduate Research Fellowship Program under Grant Number DGE-1752134.

P. D. thanks the Walmart of Yucca Valley, California for frequent and prolonged use of its wireless internet to upload hundreds of gigabytes of data. 
 

\facilities{Keck:I (HIRES), Automated Planet Finder (Levy), TESS, Gemini:Gillett (`Alopeke)}, LCOGT\\

\vspace{5mm}
\software{   
                \texttt{batman} \citep{Kreidberg2015b}, \\
                \texttt{EXOFASTv2} \citep{Eastman2013,Eastman2017,Eastman2019}, 
                \texttt{lightkurve} \citep{Lightkurve2018}, 
                \texttt{RadVel} \citep{Fulton2018}
                \texttt{SpecMatch} \citep{Petigura2015,Petigura2017b}, 
                \texttt{AstroImageJ} \citep{Collins2017}
                \texttt{LcTools} \citep{Schmitt2019}
                } \\ 

\appendix

\section{Ground-based Telescope Light Curves of \host}\label{app:ground}

In the following sections, we briefly describe each telescope that contributed to the ground-based observing campaign. Additional information summarizing the campaign is provided in Table~\ref{tab:ground_obs}.

\subsubsection{Maury Lewin Astronomical Observatory (MLO)}

The Maury Lewin Astronomical Observatory (MLO) consists of a 0.356~m Schmidt-Cassegrain telescope located near Glendora, California, USA. MLO has a SBIG STF8300M detector with a 23\arcmin x17\arcmin\ field of view. Observations of \host\ were conducted in B and I bands with various exposure times between 15 and 50~s. The data were reduced and analyzed with \texttt{AstroImageJ} (\texttt{AIJ}) following the standard differential aperture photometry protocol described by \citet{Collins2017}. 

\subsubsection{Kotizarovci Observatory (SCT)}

Kotizarovci Observatory (SCT) consists of a 0.3~m Schmidt-Cassegrain telescope located near Viskovo, Croatia. SCT has a SBIG ST7XME detector with a 15$\farcm$3 x10$\farcm$2 field of view. Observations of \host\ were conducted in a TESS-like filter with 20~s exposure times. The data were reduced and analyzed with \texttt{AIJ} following the standard differential aperture photometry protocol described by \citet{Collins2017}.  

\subsubsection{Grand-Pra Observatory (RCO)}

Grand-Pra Observatory (RCO) consists of a 0.4~m Ritchey-Chretien telescope located near Sion, Valais Switzerland. RCO has a ProLine FLI 4710 detector with a 12$\farcm$9x12$\farcm$55 field of view. Observations of \host\ were conducted in z$_{\rm s}$ filter with 15~s exposure times. The data were reduced and analyzed with \texttt{AIJ} following the standard differential aperture photometry protocol described by \citet{Collins2017}.

\subsubsection{Boyce-Astro Research Observatory (BARO)}

The Boyce-Astro Research Observatory (BARO) consists of a 0.43~m Corrected Dall-Kirkham telescope located near San Diego, California, USA. BARO has a ProLine FLI 4710 detector with a 15.6\arcmin\ square field of view. Observations of \host\ were conducted in the i$^{\prime}$ filter with 3~s exposure times. The data were reduced and analyzed with AIJ following the standard differential aperture photometry protocol described by \citet{Collins2017}.

\subsubsection{LCOGT-Haleakal\=a Observatory (LCOGT-HAL)}

The Las Cumbres Observatory Global Telescope Network \citep[LCOGT;][]{Brown2013} hosts a 0.4~m Ritchey-Chretien Cassegrain telescope on Mt. Haleakal\=a in Maui, Hawai`i, USA (LCOGT-HAL). LCOGT-HAL has a SBIG STX6303 detector with a 29$\farcm$2x19$\farcm$5 field of view. Observations of \host\ were conducted with B, z$_{\rm s}$, and i$^{\prime}$ filters with 30~s, 60~s, and 30~s exposure times, respectively. The data were reduced and analyzed with \texttt{AIJ} following the standard differential aperture photometry protocol described by \citet{Collins2017}.

\subsubsection{LCOGT-McDonald Observatory (LCOGT-McD)}

LCOGT hosts a 1.0~m Ritchey-Chretien Cassegrain telescope at McDonald Observatory near Fort Davis, Texas, USA (LCOGT-McD). LCOGT-McD has a Sinistro detector with a 26$\farcm$5 square field of view. Observations of \host\ were conducted with the z$_{\rm s}$ filter with 20~s exposure times. The data were reduced and analyzed with \texttt{AIJ} following the standard differential aperture photometry protocol described by \citet{Collins2017}.

\subsubsection{Wendelstein Observatory (W43)}

Wendelstein Observatory (W43) consists of a 0.43~m Corrected Dall-Kirkham telescope located near Bayrischzell, Germany. W43 has a SBIG STX-16803 detector with a 45\arcmin\ square field of view. Observations of \host\ were conducted in i$^{\prime}$ filter with 5~s exposure times. The data were reduced and analyzed with a custom differential aperture photometry pipeline that maximizes light curve precision by using multiple reference stars and testing various aperture sizes \citep{Dalba2016,Dalba2017a}. 

\subsubsection{Saint-Pierre-du-Mont Observatory (OPM)}

Saint-Pierre-du-Mont Observatory (OPM) consists of a 0.2~m Ritchey-Chretien telescope located near Saint-Pierre-du-Mont, France. OPM has a Atik 383 L+ detector with a 38\arcmin x29\arcmin\ field of view. Observations of \host\ were conducted in I band with 25~s or 40~s exposure times. The data were reduced and analyzed with \texttt{AIJ} following the standard differential aperture photometry protocol described by \citet{Collins2017}.

\subsubsection{LCOGT-Teide Observatory (LCOGT-TFN)}

LCOGT hosts a 0.4~m Ritchey-Chretien Cassegrain telescope at Teide Observatory in Tenerife, Spain (LCOGT-TFN). LCOGT-TFN has a SBIG STX6303 detector with a 29$\farcm$2x19$\farcm$5 field of view. Observations of \host\ were conducted with the i$^{\prime}$ filter with 30~s exposure times. The data were reduced and analyzed with \texttt{AIJ} following the standard differential aperture photometry protocol described by \citet{Collins2017}.

\subsubsection{Dragonfly Telephoto Array (DRA)}

The Dragonfly Telephoto Array (DRA), housed at the New Mexico Skies telescope hosting facility, is a remote telescope consisting of an array of small telephoto lenses roughly equivalent to a 1.0 m refractor \citep{Danieli2020}. The site is located near Mayhill, New Mexico, USA. DRA has a SBIG STF-8300M detector with a 156\arcmin x114\arcmin\ field of view. Simultaneous observations of \host\ were conducted in g$^{\prime}$ and r$^{\prime}$ bands with 15~s exposure times. The data were reduced and analyzed with a custom differential aperture photometry pipeline designed for multi-image processing and analysis.

\subsubsection{eVscope Portable Observatories (eV-A, eV-B)}

\host\ was observed with two Unistellar eVscope telescopes positioned near Joshua Tree, California, USA. The eVscope is a digital, Newtonian-like 0.114~m telescope that contains a CMOS low-light IMX224 detector with a 37\arcmin x28\arcmin\ field of view \citep{Marchis2020}. Both eVscopes observed \host\ without a filter (i.e., Clear) and with 3.975-s exposures that were subsequently stacked by a factor of 30. The stacked images were analyzed with a custom differential aperture photometry pipeline that maximizes light curve precision by using multiple reference stars and testing various aperture sizes \citep{Dalba2016,Dalba2017a}. 

\subsubsection{Indian Astronomical Observatory (HCT)}

The Indian Astronomical Observatory hosts the 2.0~m Himalayan Chandra Telescope (HCT) near Ladakh, India. HCT has a E2V detector with a 30\arcmin square field of view. Observations of \host\ were conducted in R band with 15~s exposure times. The data were reduced and analyzed with a custom differential aperture photometry pipeline that maximizes light curve precision by using multiple reference stars and testing various aperture sizes \citep{Dalba2016,Dalba2017a}. 

\subsubsection{Vainu Bappu Observatory (JCB)}

The Vainu Bappu Observatory hosts the 1.3~m J.C. Bhattacharyya Telescope (JCB) near Tamil Nadu, India. JCB has a UKATC detector with a 20\arcmin x10\arcmin\ field of view. Observations of \host\ were conducted in R band with 30~s or 50~s exposure times. The data were reduced and analyzed with a custom differential aperture photometry pipeline that maximizes light curve precision by using multiple reference stars and testing various aperture sizes \citep{Dalba2016,Dalba2017a}.

\subsubsection{Acton Sky Portal (ASP)}

The Acton Sky Portal (ASP) consists of a 0.36~m Schmidt-Cassegrain telescope located near Acton, Massachusetts, USA. ASP has a  SBIG ST8-XME detector with a 24$\farcm$2x16$\farcm$2 field of view. Observations of \host\ were conducted in r$^{\prime}$ band with 10~s exposure times. The data were reduced and analyzed with \texttt{AIJ} following the standard differential aperture photometry protocol described by \citet{Collins2017}.

\begin{figure*}
    \centering
    \includegraphics[width=0.95\textwidth]{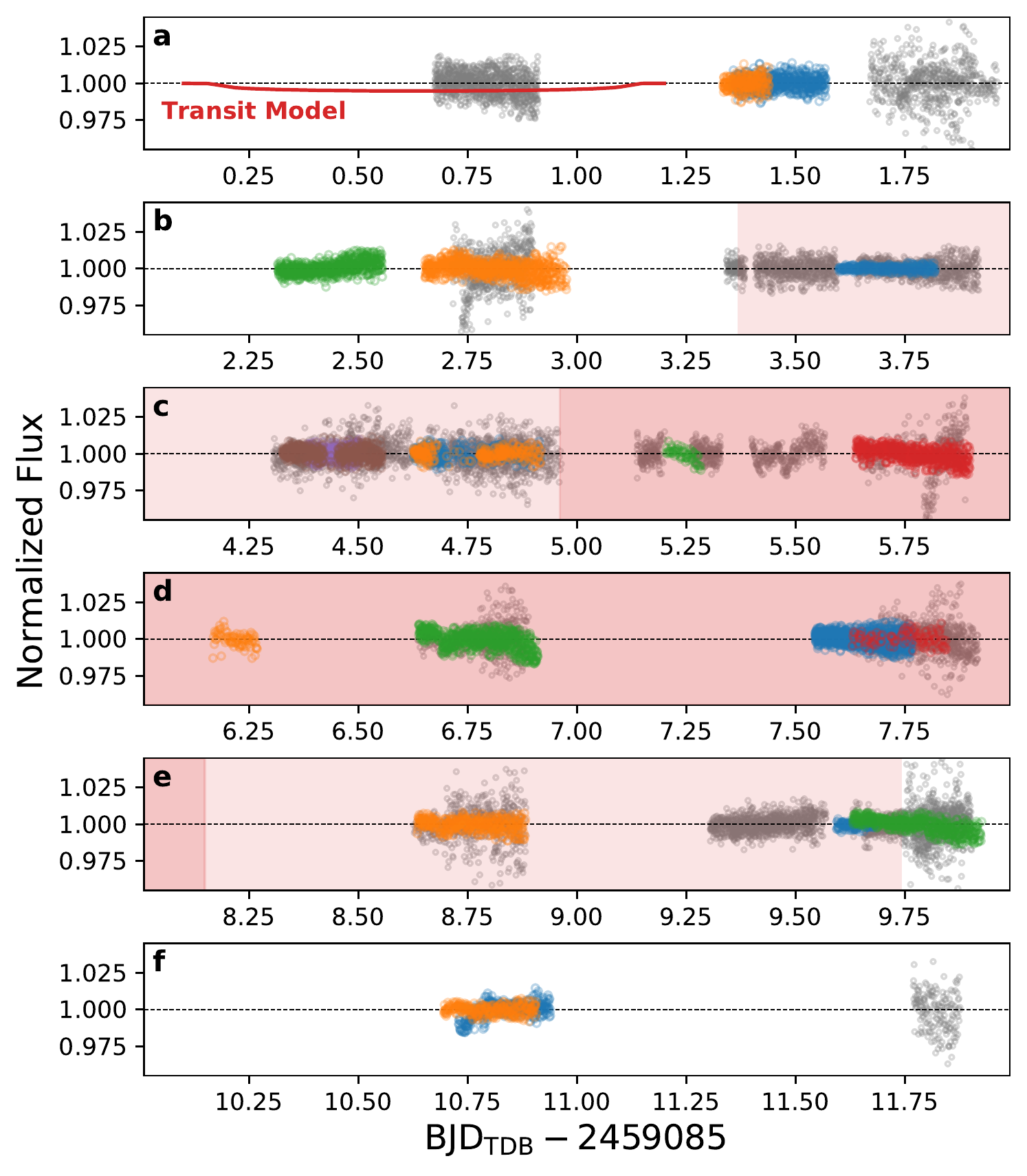}
    \caption{All ground based data acquired in the transit ephemeris refinement effort for \planet. These data have not been detrended against airmass variations. The transit model is shown in Panel a for comparison only (i.e., we are not suggesting that occurred at that time). The light and dark shaded regions denote the $2\sigma$ and $1\sigma$ transit windows, respectively, calculated from the final ephemeris in Table~\ref{tab:planet}. Data sets plotted with a color were used to refine the orbital period of \planet, whereas those in gray are shown only for completeness. These data are summarized in Table~\ref{tab:ground_obs} and will be made available upon request.}
    \label{fig:all_ground}
\end{figure*}

\clearpage
\startlongtable
\begin{deluxetable*}{cccccccccc}
\tabletypesize{\scriptsize}
\tablecaption{Summary of Ground-based Photometry of \host \label{tab:ground_obs}}
\tablehead{
  \colhead{Tel. ID} & 
  \colhead{UTC Date} &
  \colhead{Start Time$^a$} &
  \colhead{Stop Time$^a$} &
  \colhead{Filter} &
  \colhead{Exp. Time (s)} &
  \colhead{$\sigma$ (ppt)$^b$} &
  \colhead{$N_{\rm obs}\,^c$} &
  \colhead{Observer/Contact} &
  \colhead{Fig.~\ref{fig:all_ground}$^d$} }
\startdata
MLO & 2020-Aug-24 & 0.674949 & 0.910757 & B & 15 & 7.9 & 967 & Lewin & \nodata \\
SCT & 2020-Aug-24 & 1.334491 & 1.439421 & TESS & 20 & 4.5 & 255 & Srdoc & a-Orange \\
RCO & 2020-Aug-24 & 1.357160 & 1.569239 & zs & 15 & 4.7 & 548 & Girardin & a-Blue \\
BARO & 2020-Aug-25 & 1.669737 & 1.907806 & i$^{\prime}$ & 3 & 16.7 & 373 & Boyce & \nodata \\
LCOGT-HAL & 2020-Aug-25 & 1.735700 & 1.961921 & zs & 60 & 7.4 & 241 & Collins & \nodata \\
SCT & 2020-Aug-25 & 2.316263 & 2.554778 & TESS & 20 & 4.0 & 624 & Srdoc & b-Green \\
MLO & 2020-Aug-26 & 2.652399 & 2.976117 & I & 24 & 5.0 & 811 & Lewin & b-Orange \\
BARO & 2020-Aug-26 & 2.715152 & 2.899239 & i$^{\prime}$ & 3 & 12.2 & 289 & Boyce & \nodata \\
LCOGT-HAL & 2020-Aug-26 & 2.734919 & 2.896656 & B & 30 & 14.3 & 281 & Collins & \nodata \\
RCO & 2020-Aug-26 & 3.342047 & 3.593955 & zs & 15 & 5.7 & 695 & Girardin & \nodata \\
LCOGT-McD & 2020-Aug-27 & 3.599088 & 3.819390 & zs & 20 & 1.5 & 325 & Collins & b-Blue \\
MLO & 2020-Aug-27 & 3.641554 & 3.918703 & I & 24 & 5.2 & 731 & Lewin & \nodata \\
W43 & 2020-Aug-27 & 4.305405 & 4.619606 & r$^{\prime}$ & 5 & 7.6 & 550 & Steuer & \nodata \\
SCT & 2020-Aug-27 & 4.326957 & 4.554686 & TESS & 20 & 4.0 & 535 & Srdoc & c-Brown \\
OPM & 2020-Aug-27 & 4.354370 & 4.557354 & I & 25 & 8.9 & 133 & Laloum & \nodata \\
RCO & 2020-Aug-27 & 4.383018 & 4.503405 & zs & 15 & 3.6 & 364 & Girardin & c-Purple \\
LCOGT-TFN & 2020-Aug-27 & 4.397824 & 4.568805 & i$^{\prime}$ & 30 & 5.6 & 281 & Schwarz, Dragomir & \nodata \\
DRA & 2020-Aug-28 & 4.626313 & 4.917531 & r$^{\prime}$ & 15 & 3.3 & 318 & Mann & c-Orange \\
DRA & 2020-Aug-28 & 4.626314 & 4.917550 & g$^{\prime}$ & 15 & 3.6 & 427 & Mann & c-Blue \\
BARO & 2020-Aug-28 & 4.661461 & 4.894461 & i$^{\prime}$ & 3 & 11.5 & 356 & Boyce & \nodata \\
MLO & 2020-Aug-28 & 4.713041 & 4.964718 & I & 25 & 6.4 & 604 & Lewin & \nodata \\
eV-A & 2020-Aug-28 & 4.716616 & 4.878179 & Clear & 3.97 & 5.7 & 121 & Dalba & \nodata \\
eV-B & 2020-Aug-28 & 4.717048 & 4.876957 & Clear & 3.97 & 5.0 & 101 & Dalba & \nodata \\
HCT & 2020-Aug-28 & 5.137385 & 5.328852 & R & 15 & 5.4 & 359 & Unni, Thirupathi & \nodata \\
JCB & 2020-Aug-28 & 5.203377 & 5.283490 & R & 30 & 3.0 & 63 & Unni, Thirupathi & c-Green \\
LCOGT-TFN & 2020-Aug-28 & 5.399757 & 5.566928 & i$^{\prime}$ & 30 & 6.6 & 279 & Schwarz, Dragomir & \nodata \\
MLO & 2020-Aug-29 & 5.638188 & 5.898637 & I & 25 & 4.5 & 659 & Lewin & c-Red \\
eV-B & 2020-Aug-29 & 5.648009 & 5.876296 & Clear & 3.97 & 8.4 & 119 & Dalba & \nodata \\
BARO & 2020-Aug-29 & 5.656000 & 5.892000 & i$^{\prime}$ & 3 & 12.9 & 367 & Boyce & \nodata \\
eV-A & 2020-Aug-29 & 5.662142 & 5.876826 & Clear & 3.97 & 7.3 & 112 & Dalba & \nodata \\
JCB & 2020-Aug-29 & 6.167566 & 6.270573 & R & 50 & 4.4 & 83 & Unni, Thirupathi & d-Orange \\
eV-B & 2020-Aug-30 & 6.635574 & 6.872570 & Clear & 3.97 & 6.0 & 159 & Dalba & \nodata \\
MLO & 2020-Aug-30 & 6.638204 & 6.910475 & I & 25 & 4.9 & 691 & Lewin & d-Green \\
eV-A & 2020-Aug-30 & 6.720113 & 6.872948 & Clear & 3.97 & 6.1 & 101 & Dalba & \nodata \\
BARO & 2020-Aug-30 & 6.779452 & 6.888816 & i$^{\prime}$ & 3 & 14.1 & 172 & Boyce & \nodata \\
ASP & 2020-Aug-31 & 7.544901 & 7.766374 & r$^{\prime}$ & 10 & 4.2 & 1174 & Benni & d-Blue \\
eV-B & 2020-Aug-31 & 7.629679 & 7.863857 & Clear & 3.97 & 5.4 & 148 & Dalba & \nodata \\
eV-A & 2020-Aug-31 & 7.630012 & 7.845392 & Clear & 3.97 & 4.3 & 118 & Dalba & d-Red \\
MLO & 2020-Aug-31 & 7.679966 & 7.917370 & I & 25 & 5.1 & 597 & Lewin & \nodata \\
BARO & 2020-Aug-31 & 7.692682 & 7.885739 & i$^{\prime}$ & 3 & 12.8 & 299 & Boyce & \nodata \\
eV-B & 2020-Sep-01 & 8.626611 & 8.871359 & Clear & 3.97 & 6.6 & 149 & Dalba & \nodata \\
MLO & 2020-Sep-01 & 8.637218 & 8.636640 & I & 40 & 3.8 & 422 & Lewin & e-Orange \\
eV-A & 2020-Sep-01 & 8.638064 & 8.871871 & Clear & 3.97 & 5.3 & 151 & Dalba & \nodata \\
BARO & 2020-Sep-01 & 8.698078 & 8.883357 & i$^{\prime}$ & 3 & 15.2 & 293 & Boyce & \nodata \\
SCT & 2020-Sep-01 & 9.304464 & 9.545086 & TESS & 20 & 5.0 & 649 & Srdoc & \nodata \\
OPM & 2020-Sep-01 & 9.352277 & 9.569774 & I & 40 & 5.5 & 297 & Laloum & \nodata \\
LCOGT-McD & 2020-Sep-02 & 9.594154 & 9.681363 & zs & 20 & 2.1 & 123 & Collins & e-Blue \\
MLO & 2020-Sep-02 & 9.630824 & 9.925402 & I & 40 & 4.1 & 519 & Lewin & e-Green \\
DRA & 2020-Sep-02 & 9.631545 & 9.887763 & r$^{\prime}$ & 15 & 6.0 & 504 & Mann & \nodata \\
DRA & 2020-Sep-02 & 9.631549 & 9.904950 & g$^{\prime}$ & 15 & 5.9 & 595 & Mann & \nodata \\
LCOGT-HAL & 2020-Sep-02 & 9.730071 & 9.900129 & i$^{\prime}$ & 30 & 7.3 & 276 & Schwarz & \nodata \\
BARO & 2020-Sep-02 & 9.752900 & 9.880042 & i$^{\prime}$ & 3 & 20.7 & 200 & Boyce & \nodata \\
MLO & 2020-Sep-04 & 10.695742 & 10.907067 & I & 50 & 3.0 & 307 & Lewin & f-Orange \\
LCOGT-HAL & 2020-Sep-03 & 10.729537 & 10.938998 & i$^{\prime}$ & 30 & 4.6 & 355 & Schwarz,Dragomir & f-Blue \\
BARO & 2020-Sep-04 & 11.769168 & 11.875975 & i$^{\prime}$ & 3 & 12.9 & 170 & Boyce & \nodata \\
\enddata
\tablenotetext{a}{Start and stop times are listed with respect to 2459085~BJD$_{\rm TDB}$.}
\tablenotetext{b}{The $\sigma$ value is the standard deviation of the flux in parts per thousand (ppt) after the airmass correction.}
\tablenotetext{c}{The number of observations in each data set ($N_{\rm obs}$) is calculated after sigma clipping.}
\tablenotetext{d}{A combination of a letter and a color indicates that this data set was used in the ephemeris refinement of \planet\ (Section~\ref{sec:teers}) and is shown in that particular color and in that particular panel of Figure~\ref{fig:all_ground}. A value of ``\nodata'' indicates that this data set was not used in the analysis.}
\end{deluxetable*}

\end{document}